\documentclass[aps,prd,eprint,numerical,showpacs,showkeys,10pt]{revtex4-1}

\usepackage{longtable}
\usepackage{amsmath}
\usepackage{amssymb}
\usepackage{amsfonts}
\usepackage{graphicx}
\usepackage{dcolumn}
\usepackage{bm}
\usepackage{hyperref}
\usepackage{color}
\hypersetup{colorlinks=false}

\begin{document}

\title[Quasibound states on the Kerr-Newman black hole]{Instability of the charged massive scalar field on the Kerr-Newman black hole spacetime}

\date{\today}

\author{H. S. Vieira}
\email{horacio.santana.vieira@hotmail.com and horacio.santana-vieira@tat.uni-tuebingen.de}
\affiliation{Theoretical Astrophysics, Institute for Astronomy and Astrophysics, University of T\"{u}bingen, 72076 T\"{u}bingen, Germany}
\author{V. B. Bezerra}
\email{valdir@fisica.ufpb.br}
\affiliation{Departamento de F\'{i}sica, Universidade Federal da Para\'{i}ba, Caixa Postal 5008, CEP 58051-970, Jo\~{a}o Pessoa, PB, Brazil}
\author{C. R. Muniz}
\email{celio.muniz@uece.br}
\affiliation{Grupo de F\'{i}sica Te\'{o}rica (GFT), Universidade Estadual do Cear\'{a}, UECE-FECLI, Iguatu, CE, Brazil}

\begin{abstract}
We investigate the quasibound states of charged massive scalar fields in the Kerr-Newman black hole spacetime by using a new approach recently developed, which uses the polynomial conditions of the Heun functions. We calculate the resonant frequencies related to the spectrum of quasibound states, as well as its corresponding angular and radial wave eigenfunctions. We also analyze the instability of the system. These results are particularized to the cases of Schwarzschild and Kerr black holes. Additionally, we compare our analytical results with the numerical ones known in the literature. Finally, we apply the obtained results to compute the characteristic times of growth and decay of bosonic particles around a supermassive black hole situated at the center of the M87 galaxy.
\end{abstract}



\preprint{Preprint submitted to EPJC}

\maketitle


%
%


%
%
\section{Introducing the Kerr-Newman black hole spacetime}\label{KNBH}
The charged generalization of the Kerr geometry \cite{PhysRevLett.11.237} was first found as a solution to the Einstein-Maxwell field equations by Newman \textit{et al.} \cite{JMathPhys.6.918}. The connection to black holes was later discovered by Boyer and Lindquist \cite{JMathPhys.8.265}, as well as the global structure of the Kerr family of gravitational fields by Carter \cite{PhysRev.174.1559}. In this section, we will review the properties of the Kerr-Newman black hole solution, on which we want to investigate the behavior of charged massive scalar fields.

The broad aims of this paper are to obtain general solutions for the charged massive Klein-Gordon equation in the Kerr-Newman black hole spacetime and thus, revisit our previous result \cite{AnnPhys.350.14}, and use these more appropriate solutions, as compared with the ones presented in \cite{AnnPhys.350.14}, to show that the resonant frequencies describing quasibound states may be calculated analytically by using the Vieira-Bezerra-Kokkotas (VBK) approach \cite{AnnPhys.373.28,PhysRevD.104.024035}.

The line element describing a black hole with mass $M$, charge $Q$, and angular momentum per unit mass $a=J/M$ can be written as \cite{MTW:1973}
\begin{equation}
ds^{2}=g_{\mu\nu}dx^{\mu}dx^{\nu}=\frac{\Delta}{\rho^{2}}(dt-a\sin^{2}\theta\ d\phi)^{2}-\frac{\rho^2}{\Delta}dr^{2}-\rho^{2}\ d\theta^{2}-\frac{\sin^2\theta}{\rho^2}[(r^2+a^2)d\phi-a\ dt]^{2},
\label{eq:metric_KNBH}
\end{equation}
where $\rho^{2}=r^{2}+a^{2}\cos^{2}\theta$, while the function $\Delta=\Delta(r)$ has the form
\begin{equation}
\Delta=r^{2}-2Mr+a^{2}+Q^{2}.
\label{eq:Delta_KNBH}
\end{equation}
Here, $M$ is the total mass centered at the origin of the system of coordinates. Note that at the limits $a \rightarrow 0$, $Q \rightarrow 0$, and $a,Q \rightarrow 0$ the well-known Reissner-Nordstr\"{o}m, Kerr, and Schwarzschild black holes are recovered, respectively. The electromagnetic 4-vector of the black hole is given by \cite{ChinAstronAstrophys.5.365}
\begin{equation}
A_{\mu}=\frac{Qr}{\rho^{2}}(1,0,0,-a\sin^{2}\theta).
\label{eq:electromagnetic_KNBH}
\end{equation}

The horizons (null surfaces) of the Kerr-Newman black hole (KNBH) are given by the zeros of $\Delta(r)=0$, that is,
\begin{equation}
\Delta(r)=(r-r_{+})(r-r_{-})=0.
\label{eq:surface_KNBH}
\end{equation}
The solutions of this horizon surface equation, for $M^{2} > a^{2}+Q^{2}$, correspond to two horizons, namely, the interior (Cauchy) and exterior (event) horizons, which are given by $r_{-}=M-\sqrt{M^{2}-a^{2}-Q^{2}}$ and $r_{+}=M+\sqrt{M^{2}-a^{2}-Q^{2}}$, respectively. Note that when $M^{2}=a^{2}+Q^{2}$, we get an extreme Kerr-Newman black hole, that is, the interior and exterior horizons coincide at $r_{\rm e}=M$. In this work we will focus on the non-extreme case. The exterior event horizon $r_{+}$ is the outermost marginally trapped surface for the outgoing photons. In fact, this is the last surface from which the light waves could still escape from the black hole. Thus, it is meaningful to study quantum scalar particles that propagate outside the exterior event horizon, whose equation of motion is discussed in the next section.

The outline of this paper is as follows. In Sec. \ref{KG_equation}, we solve the Klein-Gordon equation in the background under consideration. In Sec. \ref{Quasibound_states}, we impose the appropriately boundary conditions and then we obtain the Hawking radiation spectrum and the resonant frequencies related to the quasibound states. In Sec. \ref{Wave_functions}, we provide the eigenfunctions by using some properties of the confluent Heun functions. In Sec. \ref{supermassive}, we apply our results to the supermassive black hole located at the center of M87 galaxy. Finally, in Sec. \ref{Final_remarks}, the conclusions are given. Here we adopted the natural units where $G \equiv c \equiv \hbar \equiv 1$.
%
%
\section{Klein-Gordon equation}\label{KG_equation}
The equation of motion which describes charged massive scalar particles propagating in a curved spacetime and in the presence of an electromagnetic field is given by
\begin{equation}
\biggl[\frac{1}{\sqrt{-g}}\partial_{\mu}(g^{\mu\nu}\sqrt{-g}\partial_{\nu})+ie(\partial_{\mu}A^{\mu})+2ieA^{\mu}\partial_{\mu}+\frac{ie}{\sqrt{-g}}A^{\mu}(\partial_{\mu}\sqrt{-g})-e^{2}A^{\mu}A_{\mu}+\mu^{2}\biggr]\Psi=0,
\label{eq:KG_equation}
\end{equation}
where $g \equiv \mbox{det}(g_{\mu\nu})$, $\Psi=\Psi(t,r,\theta,\phi)$ is the scalar wave function, and $\mu$ and $e$ are the mass and the charge of the scalar particle, respectively.

The equation of motion (\ref{eq:KG_equation}), for the metric given by Eq.~(\ref{eq:metric_KNBH}), lead in a system of two separated ordinary differential equations (ODEs) for the angular and radial parts of the scalar wave function, which are given by Eqs.~(24) and (25) in Ref.~\cite{AnnPhys.350.14}. However, in the present work, we will obtain the general solutions for both angular and radial part of the Klein-Gordon equation. Thus, with this aim, we write the scalar wave function as $\Psi(t,r,\theta,\phi)=\mbox{e}^{-i \omega t}R(r)P(\theta)\mbox{e}^{i m \phi}$, where $\omega$ is the frequency (energy) of the scalar particle, $R(r)$ is the radial function, $P(\theta)$ is the angular function, and $m$ is an integer (which can be, without loss of generality, called as the magnetic quantum number). The new ODEs for the angular and radial parts are given by
\begin{equation}
\frac{1}{\sin\theta}\frac{d}{d\theta}\biggl[\sin\theta\frac{dP(\theta)}{d\theta}\biggr]+\biggl[\lambda+a^{2}(\mu^{2}-\omega^{2})\sin^{2}\theta-\frac{m^{2}}{\sin^{2}\theta}\biggr]P(\theta)=0
\label{eq:angular_KNBH}
\end{equation}
and
\begin{equation}
\frac{d}{dr}\biggl[\Delta(r)\frac{dR(r)}{dr}\biggr]+\biggl\{\frac{[\omega(r^{2}+a^{2})-am-eQr]^{2}}{\Delta(r)}-[\lambda+\mu^{2}(r^{2}+a^{2})-2am\omega]\biggr\}R(r)=0,
\label{eq:radial_KNBH}
\end{equation}
where $\lambda$ is a separation constant to be determined. In what follows, we will provide a general solution for both angular and radial parts of the Klein-Gordon equation.
%
%
\subsection{Angular equation}
For the angular part of the Klein-Gordon equation, we define a new angular coordinate $z$ as
\begin{equation}
z=\cos\theta.
\label{eq:angular_coordinate_KNBH}
\end{equation}
Thus, by substituting Eq.~(\ref{eq:angular_coordinate_KNBH}) into Eq.~(\ref{eq:angular_KNBH}), we get
\begin{equation}
(1-z^{2})\frac{d^{2}P(z)}{dz^{2}}-2z\frac{dP(z)}{dz}+\biggl[\lambda+\gamma^{2}(1-z^{2})-\frac{m^{2}}{1-z^{2}}\biggr]P(z)=0,
\label{eq:angular_z_KNBH}
\end{equation}
where $\gamma=a\sqrt{\mu^{2}-\omega^{2}}$. This ODE has two regular singularities at $z=+1$ $(\theta=0)$ and $z=-1$ $(\theta=\pi)$ with exponents $-m/2$ and $+m/2$, respectively, as well as an irregular singularity at $z=\infty$ with rank 1. In the general case when $\lambda=\lambda(\nu)$ with arbitrary (general) degree $\nu$ ($\in \mathbb{C}$) and order $m \geq 0$ ($\in \mathbb{Z}$), Eq.~(\ref{eq:angular_z_KNBH}) is the spheroidal differential equation \cite{Erdelyi:I,Erdelyi:II,Erdelyi:III}. On the other hand, in the special case when $m$ is a nonnegative integer, and $\nu=n$ is an integer such that $n=m,m+1,m+2,\ldots$, this ODE is the so-called spheroidal wave equation. Furthermore, in the particular case when $\gamma=0$, it reduces to the well known associated Legendre equation.

Therefore, a general solution for the angular part of the Klein-Gordon equation in the KNBH spacetime is given by
\begin{equation}
P(z)=P_{\nu}^{m}(z)=C_{1}\ Ps_{\nu}^{m}(z,\gamma) + C_{2}\ Qs_{\nu}^{m}(z,\gamma),
\label{eq:analytical_solution_angular_KNBH}
\end{equation}
where $C_{1}$ and $C_{2}$ are constants to be determined, $Ps_{\nu}^{m}(z,\gamma)$ is the spheroidal function of the first kind with complex argument, and $Qs_{\nu}^{m}(z,\gamma)$ is the spheroidal function of the second kind with complex argument.

In addition, it is worth emphasizing that the angular solution given by Eq.~(\ref{eq:analytical_solution_angular_KNBH}) is regular in the whole range $0 \leq \theta \leq \pi$, for any value of the parameter $\nu$, which means that the separation constant is such that $\lambda \in \mathbb{C}$ when $\nu \in \mathbb{C}$.

In fact, these results generalize the ones obtained in our paper \cite{AnnPhys.350.14}, since one can now choose the constants $C_{1}$ and $C_{2}$ according some specific boundary conditions. The same procedure can be done for the radial part, as follows.
%
%
\subsection{Radial equation}
Now, we will also provide an analytic and general solution for the radial part of the Klein-Gordon equation in the KNBH spacetime (without the assumption of specific boundary conditions). To do this, we use Eq.~(\ref{eq:surface_KNBH}), which provides the values for the horizons, in order to write the radial equation (\ref{eq:radial_KNBH}) as
\begin{eqnarray}
&& \frac{d^{2}R(r)}{dr^{2}}+\biggl(\frac{1}{r-r_{+}}+\frac{1}{r-r_{-}}\biggr)\frac{dR(r)}{dr}\nonumber\\
&& +\frac{[\omega(r^{2}+a^{2})-am-eQr]^{2}-(r-r_{+})(r-r_{-})[\lambda+\mu^{2}(r^{2}+a^{2})-2am\omega]}{(r-r_{+})^2 (r-r_{-})^2}R(r)=0.
\label{eq:radial_2_KNBH}
\end{eqnarray}
This new form for the radial equation indicates the existence of two finite regular singularities associated with the two event horizons, $r_{+}$ and $r_{-}$, and an irregular singularity at infinity, which implies that Eq.~(\ref{eq:radial_2_KNBH}) is a Heun-type equation. Let us transform this radial equation to a more suitable form, by defining a new radial coordinate $x$, which is related to $r_{+}$ and $r_{-}$, as
\begin{equation}
x=\frac{r-r_{-}}{r_{+}-r_{-}}.
\label{eq:radial_coordinate_KNBH}
\end{equation}
This definition moves (by a homography) the two singularities $r=(r_{-},r_{+})$ to the points $x=(0,1)$, and it maintains the irregular singularity at infinity. Thus, by substituting Eq.~(\ref{eq:radial_coordinate_KNBH}) into Eq.~(\ref{eq:radial_2_KNBH}), we obtain
\begin{equation}
\frac{d^{2}R(x)}{dx^{2}}+\biggl(\frac{1}{x}+\frac{1}{x-1}\biggr)\frac{dR(x)}{dx}+\biggl[\frac{F_{0}}{x}+\frac{F_{1}}{x-1}+\frac{E_{0}}{x^{2}}+\frac{E_{1}}{(x-1)^{2}}+E_{2}\biggr]R(x)=0,
\label{eq:radial_x_KNBH}
\end{equation}
where the coefficients $E_{0,1,2}$ and $F_{0,1}$ are given by
\begin{eqnarray}
E_{0} & = & \frac{[\omega(r_{-}^{2}+a^{2})-am-eQr_{-}]^{2}}{(r_{+}-r_{-})^2},\label{eq:E0_KNBH}\\
E_{1} & = & \frac{[\omega(r_{+}^{2}+a^{2})-am-eQr_{+}]^{2}}{(r_{+}-r_{-})^2},\label{eq:E1_KNBH}\\
E_{2} & = & -(r_{+}-r_{-})^2(\mu^{2}-\omega^{2}),\label{eq:E2_KNBH}\\
F_{0} & = & \frac{1}{(r_{+}-r_{-})^2}\{a^2 [2 m^2+\mu ^2 (r_{+}-r_{-})^2]+2 a m Q \epsilon  (r_{+}+r_{-})+2 Q^2 r_{+} r_{-} \epsilon ^2+(r_{+}-r_{-})^2 (\lambda +\mu ^2 r_{-}^2)\nonumber\\
			&		& -2 \omega (2 a^3 m+a^2 Q r_{+} \epsilon +a^2 Q r_{-} \epsilon +a m r_{+}^2+a m r_{-}^2+3 Q r_{+} r_{-}^2 \epsilon -Q r_{-}^3 \epsilon)\nonumber\\
			&		& +2 \omega ^2 (a^2+r_{-}^2) (a^2+2 r_{+} r_{-}-r_{-}^2)\},\label{eq:F0_KNBH}\\
F_{1} & = & -(r_{+}-r_{-})[2 e Q \omega+(r_{+}+r_{-})(\mu^2-2\omega^2)]-F_{0}.\label{eq:F1_KNBH}
\end{eqnarray}

The radial equation (\ref{eq:radial_x_KNBH}) is almost similar to the confluent Heun equation, whose canonical form is given by \cite{Ronveaux:1995,JPhysAMathTheor.43.035203}
\begin{equation}
\frac{d^{2}U(x)}{dx^{2}}+\biggl(\alpha+\frac{\beta+1}{x}+\frac{\gamma+1}{x-1}\biggr)\frac{dU(x)}{dx}+\biggl(\frac{\xi}{x}+\frac{\zeta}{x-1}\biggr)U(x)=0,
\label{eq:confluent_Heun}
\end{equation}
where $U(x)=\mbox{HeunC}(\alpha,\beta,\gamma,\delta,\eta;x)$ are the confluent Heun functions, with the parameters $\alpha$, $\beta$, $\gamma$, $\delta$, and $\eta$ related to $\xi$, and $\zeta$ by the following expressions
\begin{eqnarray}
\xi & = & \frac{1}{2}(\alpha-\beta-\gamma+\alpha\beta-\beta\gamma)-\eta,\label{eq:sigma0_confluent_Heun}\\
\zeta & = & \frac{1}{2}(\alpha+\beta+\gamma+\alpha\gamma+\beta\gamma)+\delta+\eta.\label{eq:sigma1_confluent_Heun}
\end{eqnarray}
This standard confluent Heun function is defined via the convergent Taylor series expansion in the disc $|x| < 1$,
\begin{equation}
\mbox{HeunC}(\alpha,\beta,\gamma,\delta,\eta;x)=\sum_{n=0}^{\infty}c_{n}(\alpha,\beta,\gamma,\delta,\eta)\ x^{n},
\label{eq:HeunC_Taylor_series}
\end{equation}
where it is assumed the normalization $\mbox{HeunC}(\alpha,\beta,\gamma,\delta,\eta;0)=1$. The coefficients $c_{n}$ are given by
\begin{equation}
P_{n}c_{n}=T_{n}c_{n-1}+X_{n}c_{n-2},
\label{eq:cn}
\end{equation}
with the initial conditions $c_{-1}=0$ and $c_{0}=1$. The expressions for $P_{n}$, $T_{n}$, and $X_{n}$ are given by
\begin{eqnarray}
P_{n} & = & 1+\frac{\beta}{n},\label{eq:Pn}\\
T_{n} & = & 1+\frac{-\alpha+\beta+\gamma-1}{n}+\frac{\eta-(-\alpha+\beta+\gamma)/2-\alpha\beta/2+\beta\gamma/2}{n^{2}},\label{eq:Tn}\\
X_{n} & = & \frac{\alpha}{n^{2}}\biggl(\frac{\delta}{\alpha}+\frac{\beta+\gamma}{2}+n-1\biggr).\label{eq:Xn}
\end{eqnarray}

Now, we just need to reduce the power of the terms $1/x^{2}$ and $1/(x-1)^{2}$, as well as eliminate the term $E_{2}$ from the right hand side. In order to do this, we perform a \textit{s-homotopic transformation} of the dependent variable $R(x) \mapsto U(x)$, such that
\begin{equation}
R(x)=x^{D_{0}}(x-1)^{D_{1}}\mbox{e}^{D_{2}x}U(x),
\label{eq:s-homotopic_KNBH}
\end{equation}
where the coefficients $D_{0}$, $D_{1}$, and $D_{2}$ are the exponents of the three singular points $x=(0,1,\infty)$ in Eq.~(\ref{eq:radial_x_KNBH}). They obey to the following indicial equation
\begin{eqnarray}
s(s-1)+s+D_{0}=s^{2}+D_{0} & = & 0,\label{eq:indicial_D0_equation}\\
s^{2}+D_{1} & = & 0,\label{eq:indicial_D1_equation}\\
s^{2}+D_{2} & = & 0,\label{eq:indicial_D2_equation}
\end{eqnarray}
whose roots are given by
\begin{eqnarray}
s_{\pm}^{x=0} & = & \pm i\frac{\omega(r_{-}^{2}+a^{2})-am-eQr_{-}}{r_{+}-r_{-}} \equiv D_{0},\label{eq:D0_KNBH}\\
s_{\pm}^{x=1} & = & \pm i\frac{\omega(r_{+}^{2}+a^{2})-am-eQr_{+}}{r_{+}-r_{-}} \equiv D_{1},\label{eq:D1_KNBH}\\
s_{\pm}^{x=\infty} & = & \pm (r_{+}-r_{-}) \sqrt{\mu^{2}-\omega^{2}} \equiv D_{2}.\label{eq:D2_KNBH}
\end{eqnarray}
Thus, by substituting Eqs.~(\ref{eq:s-homotopic_KNBH})-(\ref{eq:D2_KNBH}) into Eq.~(\ref{eq:radial_x_KNBH}), we derive a new equation for the radial function $U(x)$, namely,
\begin{equation}
\frac{d^{2}U(x)}{dx^{2}}+\biggl(2D_{2}+\frac{1+2D_{0}}{x}+\frac{1+2D_{1}}{x-1}\biggr)\frac{dU(x)}{dx}+\biggl(\frac{\xi}{x}+\frac{\zeta}{x-1}\biggr)U(x)=0,
\label{eq:radial_3_KNBH}
\end{equation}
where
\begin{eqnarray}
\xi & = & -D_{0}-D_{1}-2D_{0}D_{1}+D_{2}+2D_{0}D_{2}+F_{0},\label{eq:sigma0_radial_KNBH}\\
\zeta & = & D_{0}+D_{1}+2D_{0}D_{1}+D_{2}+2D_{1}D_{2}+F_{1}.\label{eq:sigma1_radial_KNBH}
\end{eqnarray}

Therefore, we can see that Eq.~(\ref{eq:radial_3_KNBH}) has the form of a confluent Heun equation (\ref{eq:confluent_Heun}). As a conclusion, we can write a general solution for the radial part of the Klein-Gordon equation in the KNBH spacetime as
\begin{equation}
R_{j}(x)=x^{D_{0}}(x-1)^{D_{1}}\mbox{e}^{D_{2}x}\{C_{1,j}\ U_{1,j}(x)+C_{2,j}\ U_{2,j}(x)\},
\label{eq:analytical_solution_radial_KNBH}
\end{equation}
where $C_{1,j}$ and $C_{2,j}$ are constants to be determined, and $j=\{0,1,\infty\}$ labels the solutions at each singular point. Thus,
the pair of linearly independent solutions at $x=0$ ($r=r_{-}$) is given by
\begin{eqnarray}
U_{1,0}(x) & = & \mbox{HeunC}(\alpha,\beta,\gamma,\delta,\eta;x),\label{eq:U10}\\
U_{2,0}(x) & = & x^{-\beta}\mbox{HeunC}(\alpha,-\beta,\gamma,\delta,\eta;x),\label{eq:U20}
\end{eqnarray}
where the parameters $\alpha$, $\beta$, $\gamma$, $\delta$, and $\eta$ are given by
\begin{eqnarray}
\alpha	& = & 2D_{2},\label{eq:alpha0_radial_KNBH}\\
\beta		& = & 2D_{0},\label{eq:beta0_radial_KNBH}\\
\gamma	& = & 2D_{1},\label{eq:gamma0_radial_KNBH}\\
\delta	& = & F_{0}+F_{1},\label{eq:delta0_radial_KNBH}\\
\eta		& = & -F_{0}.\label{eq:eta0_radial_KNBH}
\end{eqnarray}
Similarly, the pair of linearly independent solutions at $x=1$ ($r=r_{+}$) is given by
\begin{eqnarray}
U_{1,1}(x) & = & \mbox{HeunC}(\alpha,\beta,\gamma,\delta,\eta;1-x),\label{eq:U11}\\
U_{2,1}(x) & = & (1-x)^{-\beta}\mbox{HeunC}(\alpha,-\beta,\gamma,\delta,\eta;1-x),\label{eq:U21}
\end{eqnarray}
where the parameters $\alpha$, $\beta$, $\gamma$, $\delta$, and $\eta$ are given by
\begin{eqnarray}
\alpha	& = & 2D_{2},\label{eq:alpha1_radial_KNBH}\\
\beta		& = & 2D_{1},\label{eq:beta1_radial_KNBH}\\
\gamma	& = & 2D_{0},\label{eq:gamma1_radial_KNBH}\\
\delta	& = & -(F_{0}+F_{1}),\label{eq:delta1_radial_KNBH}\\
\eta		& = & F_{1}.\label{eq:eta1_radial_KNBH}
\end{eqnarray}
Finally, the two linearly independent solutions of the confluent Heun equation at $x=\infty$ ($r=\infty$) can be expanded (in a sector) in the following asymptotic series
\begin{eqnarray}
U_{1,\infty}(x) & = & x^{-(\frac{\beta+\gamma+2}{2}+\frac{\delta}{\alpha})},\label{eq:U1i}\\
U_{2,\infty}(x) & = & x^{-(\frac{\beta+\gamma+2}{2}-\frac{\delta}{\alpha})}\mbox{e}^{- \alpha x}.\label{eq:U2i}
\end{eqnarray}

Note that the final expressions for these parameters will depend on the signs to be chosen for the coefficients $D_{0,1,2}$, which are given by Eqs.~(\ref{eq:D0_KNBH})-(\ref{eq:D2_KNBH}); the positive sign corresponds to the solution obtained in \cite{AnnPhys.350.14}. For our purposes in this work, the negative sign is the correct choice for these coefficients, which will be justified in the discussion of the quasibound states. Therefore, from now on we will consider the following expressions for the coefficients $D_{0,1,2}$:
\begin{eqnarray}
D_{0} & = & -i\frac{\omega(r_{-}^{2}+a^{2})-am-eQr_{-}}{r_{+}-r_{-}},\label{eq:D0_QBSs_KNBH}\\
D_{1} & = & -i\frac{\omega(r_{+}^{2}+a^{2})-am-eQr_{+}}{r_{+}-r_{-}},\label{eq:D1_QBSs_KNBH}\\
D_{2} & = & -(r_{+}-r_{-})\sqrt{\mu^{2}-\omega^{2}}.\label{eq:D2_QBSs_KNBH}
\end{eqnarray}

The assumption of a specific asymptotic behavior on the aforementioned general solutions near the exterior event horizon $r=r_{+}$ and spatial infinity can lead to various physical solutions. In what follows, we will use these analytical solutions for the radial part of the Klein-Gordon equation in the KNBH spacetime, and the properties of the confluent Heun functions, to discuss the quasibound states and their corresponding wave eigenfunctions.
%
%
\section{Quasibound states}\label{Quasibound_states}
In this section, we will investigate the spectrum of quasibound states, which are solutions to the equation of motion satisfying two boundary conditions; they are ingoing waves at the exterior event horizon, and tend to zero at infinity. To do this, we need to calculate the quasispectrum of resonant frequencies for charged massive scalar in the KNBH spacetime.

Since the flux of particles crosses into the exterior horizon surface, the spectrum of quasibound states has complex frequencies, which can be expressed as $\omega=\omega_{R}+i\omega_{I}$, where $\omega_{R}=\mbox{Re}[\omega]$ and $\omega_{I}=\mbox{Im}[\omega]$ are the real and imaginary parts, respectively. The wave solution decays with the time if $\omega_{I} < 0$, otherwise, it grows if $\omega_{I} > 0$.

The common approach used to derive the characteristic resonance equation is to solve the radial equation in two different asymptotic regions and then, by using a standard matching procedure, evaluate these two solutions in their common overlap region \cite{PhysRevD.85.044031,PhysLettB.749.167,PhysRevD.103.044062}. Here, we will obtain the spectrum of quasibound states by using the technique recently developed by Vieira and Kokkotas \cite{PhysRevD.104.024035}, which corresponds to impose the appropriate boundary conditions to the radial solution and then use the polynomial condition of the Heun function as a matching procedure. In fact, this method is an extension of the one developed by Vieira and Bezerra \cite{AnnPhys.373.28} used to find general expressions for the resonant frequencies and hence we call it as the VBK approach.

The first boundary condition is to demand that the radial solution should describe an ingoing wave at the exterior event horizon. In this limit, which means $r \rightarrow r_{+}$ (or $x \rightarrow 1$), the radial solution given by Eq.~(\ref{eq:analytical_solution_radial_KNBH}) has the following asymptotic behavior
\begin{equation}
\lim_{r \rightarrow r_{+}} R_{1}(r) \sim C_{1,1}\ (r-r_{+})^{D_{1}}+C_{2,1}\ (r-r_{+})^{-D_{1}},
\label{eq:asymptotic_KNBH}
\end{equation}
where all remaining constants have been included in $C_{1,1}$ and $C_{2,1}$. Then, from Eq.~(\ref{eq:D1_QBSs_KNBH}), we get
\begin{equation}
D_{1}=-\frac{i}{2\kappa_{+}}(\omega-\omega_{c}),
\label{eq:D0_QBSs_full_KNBH}
\end{equation}
where $\omega_{c}=m\Omega_{+}+e\Phi_{+}$ is a critical value related to the superradiance phenomenon, with $\Omega_{+}=a/(r_{+}^{2}+a^{2})$ and $\Phi_{+}=Qr_{+}/(r_{+}^{2}+a^{2})$. The gravitational acceleration on the exterior event horizon $\kappa_{+}$ is given by
\begin{equation}
\kappa_{+} \equiv \frac{1}{2(r_{+}^{2}+a^{2})} \left.\frac{d\Delta(r)}{dr}\right|_{r=r_{+}} = \frac{r_{+}-r_{-}}{2(r_{+}^{2}+a^{2})}.
\label{eq:grav_acc_KNBH}
\end{equation}
Thus, Eq.~(\ref{eq:asymptotic_KNBH}) can be rewritten as
\begin{equation}
\lim_{r \rightarrow r_{+}} R_{1}(r) \sim C_{1,1}\ R_{{\rm in},1} + C_{2,1}\ R_{{\rm out},1},
\label{eq:full_wave_KNBH}
\end{equation}
where the ingoing, $R_{{\rm in},1}$, and outgoing, $R_{{\rm out},1}$, radial solutions are given by
\begin{equation}
R_{{\rm in},1}=R_{{\rm in},1}(r>r_{+})=(r-r_{+})^{-\frac{i}{2\kappa_{+}}(\omega-\omega_{c})}
\label{eq:radial_in_KNBH}
\end{equation}
and
\begin{equation}
R_{{\rm out},1}=R_{{\rm out},1}(r>r_{+})=(r-r_{+})^{+\frac{i}{2\kappa_{+}}(\omega-\omega_{c})}.
\label{eq:radial_out_KNBH}
\end{equation}
Therefore, in order to fully satisfy the first boundary condition, we must impose that $C_{2,1}=0$ in Eq.~(\ref{eq:full_wave_KNBH}), and in Eq.~(\ref{eq:analytical_solution_radial_KNBH}) as well. Thus, we get
\begin{equation}
\lim_{r \rightarrow r_{+}} R_{1}(r) \sim C_{1,1}\ R_{{\rm in},1}.
\label{eq:full_wave_2_KNBH}
\end{equation}

The second boundary condition is such that the radial solution must tend to zero far from the black hole at asymptotic infinity. In this limit, which means $r \rightarrow \infty$ (or $x \rightarrow \infty$), we will use the two linearly independent solutions of the confluent Heun equation at spatial infinity, which are given by Eqs.~(\ref{eq:U1i}) and (\ref{eq:U2i}), to obtain the following asymptotic behavior
\begin{equation}
\lim_{r \rightarrow \infty} R_{\infty}(r) \sim C_{1,\infty}\ r^{-1}\ r^{-p}\ \mbox{e}^{-q r},
\label{eq:infty_solution_radial_2_KNBH}
\end{equation}
where we have imposed that $C_{2,\infty}=0$. The parameters $p$, and $q$ are given by
\begin{eqnarray}
p & = & \frac{e Q \omega + M (\mu^{2}-2\omega^{2})}{q},\label{eq:p_infty_solution_radial_2_KNBH}\\
q & = & \sqrt{\mu^{2}-\omega^{2}}.\label{eq:q_infty_solution_radial_2_KNBH}
\end{eqnarray}
The sign of the real part of $q$ determines the behavior of the wave function as $r \rightarrow \infty$. Thus, if $\mbox{Re}[q] > 0$, the radial solution tends to zero at spatial infinity, which is the definition of the quasibound state solutions; whereas if $\mbox{Re}[q] < 0$, the solution diverges. The final behavior of the scalar wave function will be determined when we know the values of the frequency $\omega$, which depend on the values of the parameters $M$, $\mu$, $a$, $m$, $e$, and $Q$. It will be obtained in what follows.
%
%
\subsection{Polynomial condition}
Now, we will follow the VBK approach to find the quasibound states of scalar fields in the background under consideration, which uses the polynomial condition of the Heun functions as a matching procedure for the two (asymptotic) radial solutions in their common overlap region.

It is known that the confluent Heun polynomials are solutions of the confluent Heun equation valid at all its singularities, in the sense of being simultaneously a Frobenius solution at each one of them \cite{Ronveaux:1995}. In fact, this is a kind of standard matching procedure. Therefore, in order to satisfy the second boundary condition for quasibound states, it is necessary that the radial solution must be written in terms of the confluent Heun polynomials, which means that we need to derive a form of the confluent Heun functions that presents a polynomial behavior.

Thus, we will use the fact that the confluent Heun functions become polynomials of degree $N \geq 0$ if and only if they satisfy the following two conditions \cite{JPhysAMathTheor.43.035203}:
\begin{eqnarray}
\frac{\delta}{\alpha}+\frac{\beta+\gamma}{2}+N+1 & = & 0,\label{eq:delta-condition}\\
\Delta_{N+1}(\xi) & = & 0,\label{eq:Delta-condition}
\end{eqnarray}
where $\xi$ is given by Eq.~(\ref{eq:sigma0_confluent_Heun}). The first condition, given by Eq.~(\ref{eq:delta-condition}), is called $\delta$-condition, which is equivalent to $X_{N+2}=0$ in Eq.~(\ref{eq:Xn}). The second condition, given by Eq.~(\ref{eq:Delta-condition}), is called $\Delta$-condition, which is equivalent to the requirement $c_{N+1}=0$ in Eq.~(\ref{eq:cn}). Thus, $\Delta_{N+1}(\xi)$ is the three-diagonal determinant given by
\begin{equation}
\!\left|\!
\begin{array}{ccccccc}
	\xi\!-\!z_{1} & 1\!\left(\!1\!+\!\beta\!\right)\! & 0 & \!\cdots\! & 0 & 0 & 0 \\
	N\alpha & \xi\!-\!z_{2}\!+\!1\alpha & 2\!\left(\!2\!+\!\beta\!\right)\! & \!\cdots\! & 0 & 0 & 0 \\
	0 & \!\left(\!N\!-\!1\!\right)\!\alpha & \xi\!-\!z_{3}\!+\!2\alpha & \!\cdots\! & 0 & 0 & 0 \\
	\!\vdots\! & \!\vdots\! & \!\vdots\! & \!\ddots\! & \!\vdots\! & \!\vdots\! & \!\vdots\! \\
	0 & 0 & 0 & \!\cdots\! & \xi\!-\!z_{N\!-\!1}\!+\!\!\left(\!N\!-\!2\!\right)\!\alpha & \!\left(\!N\!-\!1\!\right)\!\!\left(\!N\!-\!1\!+\!\beta\!\right)\! & 0 \\
	0 & 0 & 0 & \!\cdots\! & 2\alpha & \xi\!-\!z_{N}\!+\!\!\left(\!N\!-\!1\!\right)\!\alpha & N\!\left(\!N\!+\!\beta\!\right)\! \\
	0 & 0 & 0 & \!\cdots\! & 0 & 1\alpha & \xi\!-\!z_{N\!+\!1}\!+\!N\alpha \\
\end{array}
\!\right|,
\label{eq:cond_polin_determ}
\end{equation}
where $z_{n}=(n-1)(n+\beta+\gamma)$. Here, $N=0,1,2,\ldots$ is the overtone number (which can be, without loss of generality, called as the principal quantum number). Such polynomial solutions are denoted by $\mbox{HeunCp}_{N}(x)$ and their properties will be discussed in the next section.

It is worth emphasizing that similar equations to Eqs.~(\ref{eq:delta-condition}) and (\ref{eq:Delta-condition}) have also been obtained by Ishkhanyan \textsl{et al.} \cite{AIPAdvances.4.087132} and Kraniotis \cite{ClassQuantumGrav.33.225011}, when they investigated expansions of the solutions to the confluent Heun equation in terms of the Kummer confluent hypergeometric functions and applied to examine behavior of charged massive scalar particles in the Kerr-Newman black hole spacetime, whose solution of the equation of motion is given in terms of the confluent Heun function (see for instance Eqs.~(163), (164), (194), and (195) in Ref. \cite{ClassQuantumGrav.33.225011}).

The $\delta$-condition is used to find the frequency eigenvalues. These frequency eigenvalues will not depend on the separation constant $\lambda$, since the parameter $\eta$ does not appear in this first condition. Next, the $\Delta$-condition determines the values of the separation constant $\lambda$, which must be used to find both the radial and angular wave eigenfunctions. In summary, we could obtain the eigenvalues of the separation constant $\lambda$, which corresponds to the appropriate eigenvalues of the frequencies $\omega$, from
the polynomial solution for the radial equation and then use it to show the (regular) angular behavior of the scalar quasibound states in the KNBH spacetime.

In addition, it is worth commenting that these two conditions could be applied in an inverse order, that is, depending on the quantities involved in the parameters of the confluent Heun functions, the first condition gives the restrictions on the separation constant, and the second one gives the frequency eigenvalues.

Then, by imposing the $\delta$-condition given by Eq.~(\ref{eq:delta-condition}), we obtain the following characteristic resonance equation
\begin{equation}
\frac{eQ\omega+M(\mu^2-2\omega^2)}{\sqrt{\mu ^2-\omega ^2}}+\frac{(N+1)\sqrt{M^2-a^2-Q^2}+i[am+MQe+(Q^2-2M^2)\omega]}{\sqrt{M^2-a^2-Q^2}}=0,
\label{eq:resonance_equation_KNBH}
\end{equation}
which can be rewritten as
\begin{equation}
a_{0}+a_{1}\ \omega+a_{2}\ \omega^{2}+a_{3}\ \omega^{3}+a_{4}\ \omega^{4}=0,
\label{eq:characteristic_resonance_equation_KNBH}
\end{equation}
where
\begin{eqnarray}
a_{0} & = &	\mu ^2 \{a m [2 i (N+1) \sqrt{M^2-a^2-Q^2}-2 M Q e]-a^2 [m^2-\mu ^2 M^2+(N+1)^2]\nonumber\\
			&		&	+2 i M Q e (N+1) \sqrt{M^2-a^2-Q^2}-\mu ^2 M^4+M^2 [Q^2 (\mu ^2-e ^2)+(N+1)^2]-Q^2 (N+1)^2\},\nonumber\\
a_{1} & = &	2 \mu ^2 [(i Q^2 -2 i M^2) (N+1) \sqrt{M^2-a^2-Q^2}+(a^2+M^2) M Q e +a m (2 M^2-Q^2)],\nonumber\\
a_{2} & = &	a^2 [m^2-4 \mu ^2 M^2+Q^2 e ^2+(N+1)^2]+2 a m [M Q e -i (N+1) \sqrt{M^2-a^2-Q^2}]\nonumber\\
			&		&	-2 i M Q e (N+1) \sqrt{M^2-a^2-Q^2}+Q^4 (e ^2-\mu ^2)+(Q^2-M^2) (N+1)^2,\nonumber\\
a_{3} & = & -2 [(i Q^2-2 i M^2) (N+1) \sqrt{M^2-a^2-Q^2}+(2 a^2+Q^2) M Q e +a m (2 M^2-Q^2)],\nonumber\\
a_{4} & = & 4a^2M^2+Q^4.
\label{eq:a_KNBH}
\end{eqnarray}

Now, let us do the first validation of this approach by computing the frequency spectrum in the nonrelativistic limit. To do this, we consider that the scalar field bounds to a Schwarzschild black hole; that is the low frequency limit in which $M\omega \rightarrow 0$ \cite{JCAP.01.006}. In this limit, the energy spectra of bound states, $\mathcal{E}_{\bar{N}}$, are the solutions of a second order equation in $\omega$, and are given by
\begin{equation}
\mathcal{E}_{\bar{N}} = \pm \mu c^{2}\sqrt{1-\frac{G^{2}M^{2}\mu^{2}}{\hbar^{2}c^{2}\bar{N}^{2}}},
\label{eq:hydrogenic_spectrum_1}
\end{equation}
where $\bar{N}=N+1$, and the signs $+$ and $-$ are related to the particle and antiparticle spectrum, respectively. Note that we have recovered all the universal constants. Then, for a scalar particle in the nonrelativistic approximation, which means that $\mathcal{O}(1/c^{2}) \rightarrow 0$, after subtracting rest energy, we get
\begin{equation}
\mathcal{E}_{\bar{N}} \approx -\frac{G^{2}M^{2}\mu^{3}}{2\hbar^{2}\bar{N}^{2}},
\label{eq:hydrogenic_spectrum_2}
\end{equation}
This is the same result obtained by Lasenby \textit{et al.} \cite{PhysRevD.72.105014} and Dolan \cite{PhysRevD.76.084001}.

In what follows, as a second validation of this approach, we solve the characteristic resonance equation (\ref{eq:characteristic_resonance_equation_KNBH}) for the particular cases of Schwarzschild ($a=0$, $Q=0$) and Kerr ($Q=0$) black holes, and then we compare our analytical results with the numerical ones obtained by Dolan \cite{PhysRevD.76.084001}. Next, we will present the spectrum of quasibound state frequencies in the KNBH spacetime and then we will compare our analytical results with the ones obtained by Furuhashi and Nambu \cite{ProgTheorPhys.112.983}; they used two approaches: numerical and analytic approximation.

From here on, we set $M=1$, which means that $r$ and $a$ will be measured in units of $M$, and $\mu$ and $\omega$ in units of $1/M$. Thus, our results will be given in terms of the gravitational coupling $M\mu$, which is the (dimensionless) product of the black hole mass $M$ and the scalar field mass $\mu$.
%
%
\subsection{Schwarzschild black hole}
In the particular case when $a=0$ and $Q=0$, the metric given by Eq.~(\ref{eq:metric_KNBH}) reduces to the Schwarzschild form. This implies that $a_{4}=0$ in Eq.~(\ref{eq:characteristic_resonance_equation_KNBH}), which means that the characteristic resonance equation has three solutions.

The Wolfram Mathematica$^{\mbox{\textregistered}}$ 12.3 can solve cubic equations exactly. However, the final expressions are quite long, and for this reason no insight is gained by writing them out. Thus, instead of doing this, we will discuss some of their features. The obtained quasispectrum is complex, as expected, and the eigenvalues will be denoted as $\omega_{N}^{(j)}$, where $j=1,2,3$ labels the solutions. As a consequence, the coefficient $q$, which is given by Eq.~(\ref{eq:q_infty_solution_radial_2_KNBH}), can be denoted as $q_{N}^{(j)}$.

Some values of the massive scalar resonant frequencies in the Schwarzschild black hole, as well as the corresponding coefficient $q_{N}^{(j)}$, are shown in Table \ref{tab:1_KNBH}, as functions of the gravitational coupling $M\mu$. From Table \ref{tab:1_KNBH}, we conclude that all resonant frequencies $\omega_{N}^{(j)}$ are physically admissible, since $\mbox{Re}[q_{N}^{(j)}] > 0$ and hence the radial solution, given by Eq.~(\ref{eq:analytical_solution_radial_KNBH}) with $C_{2,j}=0$, describes ingoing waves at the exterior event horizon and tends to zero at asymptotic infinity, which represents the spectrum of quasibound state frequencies for massive scalar particles in the Schwarzschild black hole spacetime. However, from a numerical (and dimensional) point of view, Eq.~(\ref{eq:resonance_equation_KNBH}) seems to have just two solutions and hence we may conclude that one of its solutions is incidental; in this scenario, the third solution $\omega_{N}^{(3)}$ is incidental, which may describe an unstable system, since there is a change in the sign of its imaginary part when $M\mu=1$. On the other hand, two solutions have the same decay rate and opposite oscillation frequency, that is, $\mbox{Im}[\omega_{N}^{(1)}]=\mbox{Im}[\omega_{N}^{(2)}]$ and $\mbox{Re}[\omega_{N}^{(1)}]=-\mbox{Re}[\omega_{N}^{(2)}]$, respectively. This may describe the pair production of particle and its antiparticle from a boson.

\begin{figure}[p]
\centering
\includegraphics[width=1\columnwidth]{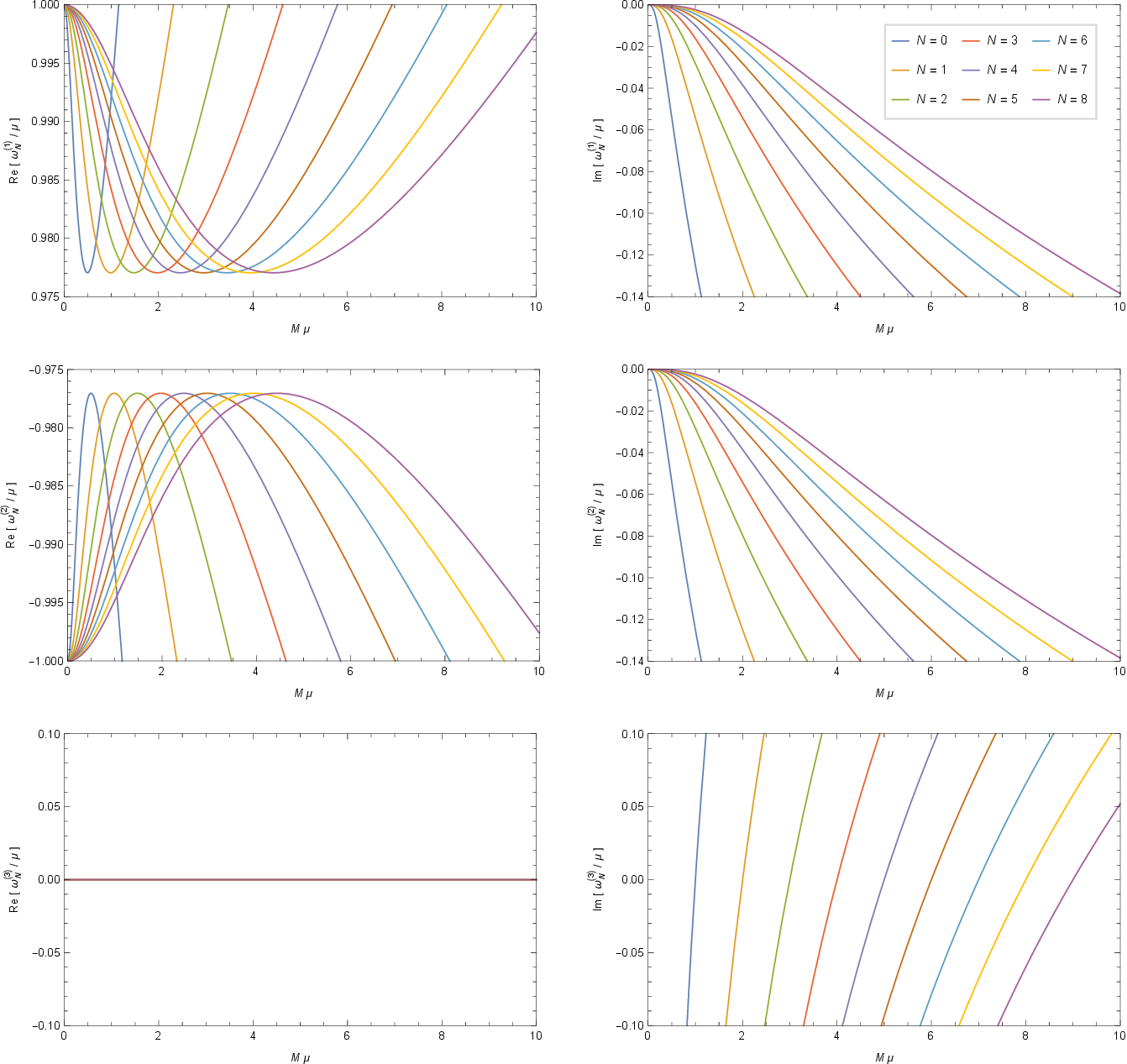}
\caption{Massive scalar resonant frequencies in the Schwarzschild black hole spacetime. The left plots show the oscillation frequency $\mbox{Re}[\omega_{N}^{(j)}/\mu]$, while the right plots show the decay (or growth) rate $\mbox{Im}[\omega_{N}^{(j)}/\mu]$.}
\label{fig:Fig1}
\end{figure}

In order to compare our analytical results with the numerical ones obtained by Dolan \cite{PhysRevD.76.084001}, we show the behavior of the massive scalar resonant frequencies $\omega_{N}^{(j)}/\mu$ in Fig.~\ref{fig:Fig1}, as a function of the gravitational coupling $M\mu$. In Fig.~\ref{fig:Fig1} we can see that the resonant frequencies $\omega_{N}^{(1)}/\mu$ have a behavior which is similar to the results presented in Fig.~3 of Dolan's paper \cite{PhysRevD.76.084001}; here, in comparison with Dolan's paper, the overtone number $N$ ``plays the role'' of the eigenvalue $l$. Therefore, it indicates that the VBK approach is an analytical generalization of the numerical approaches known in the literature (see Ref.~\cite{PhysRevD.76.084001} and references therein). Furthermore, this method provides a complete set of resonant frequencies; in this case, we get three frequency eigenvalues, which may describe the full behavior of massive scalar particles in the Schwarzschild black hole spacetime.

\begin{turnpage}
\begin{table}[ht]
\caption{Values of the resonant frequencies $\omega_{N}^{(j)}$, and the real part of the corresponding coefficient $q_{N}^{(j)}$, in the Schwarzschild black hole spacetime for $0.1 \leq M\mu \leq 1.0$. We focus on the fundamental mode $N=0$.}
\label{tab:1_KNBH}
\begin{tabular}{c|c|c|c|c|c|c}
\hline
\noalign{\smallskip}\hline\noalign{\smallskip}
$M\mu$ & $\omega_{0}^{(1)}$   & $\mbox{Re}[q_{0}^{(1)}]$ & $\omega_{0}^{(2)}$    & $\mbox{Re}[q_{0}^{(2)}]$ & $\omega_{0}^{(3)}$         & $\mbox{Re}[q_{0}^{(3)}]$ \\
\noalign{\smallskip}\hline\noalign{\smallskip}
0.1    & $0.099567-0.000172i$ & $0.009468$               & $-0.099567-0.000172i$ & $0.009468$               & $-0.249654i$               & $0.268937$               \\
0.2    & $0.197527-0.001989i$ & $0.033529$               & $-0.197527-0.001989i$ & $0.033529$               & $-0.246021i$               & $0.317059$               \\
0.3    & $0.294400-0.006947i$ & $0.065883$               & $-0.294400-0.006947i$ & $0.065883$               & $-0.236106i$               & $0.381767$               \\
0.4    & $0.391177-0.015379i$ & $0.103072$               & $-0.391177-0.015379i$ & $0.103072$               & $-0.219241i$               & $0.456143$               \\
0.5    & $0.488523-0.027094i$ & $0.143487$               & $-0.488523-0.027094i$ & $0.143487$               & $-0.195811i$               & $0.536975$               \\
0.6    & $0.586810-0.041785i$ & $0.186327$               & $-0.586810-0.041785i$ & $0.186327$               & $-0.166430i$               & $0.622655$               \\
0.7    & $0.686236-0.059156i$ & $0.231140$               & $-0.686236-0.059156i$ & $0.231140$               & $-0.131687i$               & $0.712279$               \\
0.8    & $0.786908-0.078953i$ & $0.277642$               & $-0.786908-0.078953i$ & $0.277642$               & $-0.092092i$               & $0.805283$               \\
0.9    & $0.888875-0.100962i$ & $0.325642$               & $-0.888875-0.100962i$ & $0.325642$               & $-0.048076i$               & $0.901283$               \\
1.0    & $0.992157-0.125000i$ & $0.375000$               & $-0.992157-0.125000i$ & $0.375000$               & $\ 0.000000$               & $1.000000$               \\
\noalign{\smallskip}\hline\noalign{\smallskip}
\hline
\end{tabular}
\end{table}
\end{turnpage}

%
%
\subsection{Kerr black hole}
Now, in the particular case when $Q=0$, the metric given by Eq.~(\ref{eq:metric_KNBH}) reduces to the Kerr form. This implies that the characteristic resonance equation (\ref{eq:characteristic_resonance_equation_KNBH}) has four solutions, which can also be found by using the Wolfram Mathematica$^{\mbox{\textregistered}}$ 12.3.

The obtained quasispectrum is also complex, and the eigenvalues are denoted by $\omega_{mN}^{(j)}$, where $j=1,2,3,4$ labels the solutions. As a consequence, the coefficient $q$, which is given by Eq.~(\ref{eq:q_infty_solution_radial_2_KNBH}), is denoted as $q_{mN}^{(j)}$.

Some values of the massive scalar resonant frequencies in the Kerr black hole, as well as the corresponding coefficient $q_{mN}^{(j)}$, are shown in Table \ref{tab:2_KNBH}, as functions of the gravitational coupling $M\mu$. From Table \ref{tab:2_KNBH}, we conclude that all resonant frequencies $\omega_{mN}^{(j)}$ are physically admissible, since $\mbox{Re}[q_{mN}^{(j)}] > 0$ and hence they represent the spectrum of quasibound state frequencies for massive scalar particles in the Kerr black hole spacetime. However, as Eq.~(\ref{eq:resonance_equation_KNBH}) seems to have just two solutions, we may conclude that the first $\omega_{mN}^{(1)}$ and second solutions $\omega_{mN}^{(2)}$ are incidental in this scenario. On the other hand, two solutions have the same decay rate and opposite oscillation frequency, that is, $\mbox{Im}[\omega_{N}^{(3)}]=\mbox{Im}[\omega_{N}^{(4)}]$ and $\mbox{Re}[\omega_{N}^{(3)}]=-\mbox{Re}[\omega_{N}^{(4)}]$, respectively, which may describe the pair production of particle and its antiparticle from a boson.

\begin{figure}[p]
\centering
\includegraphics[width=1\columnwidth]{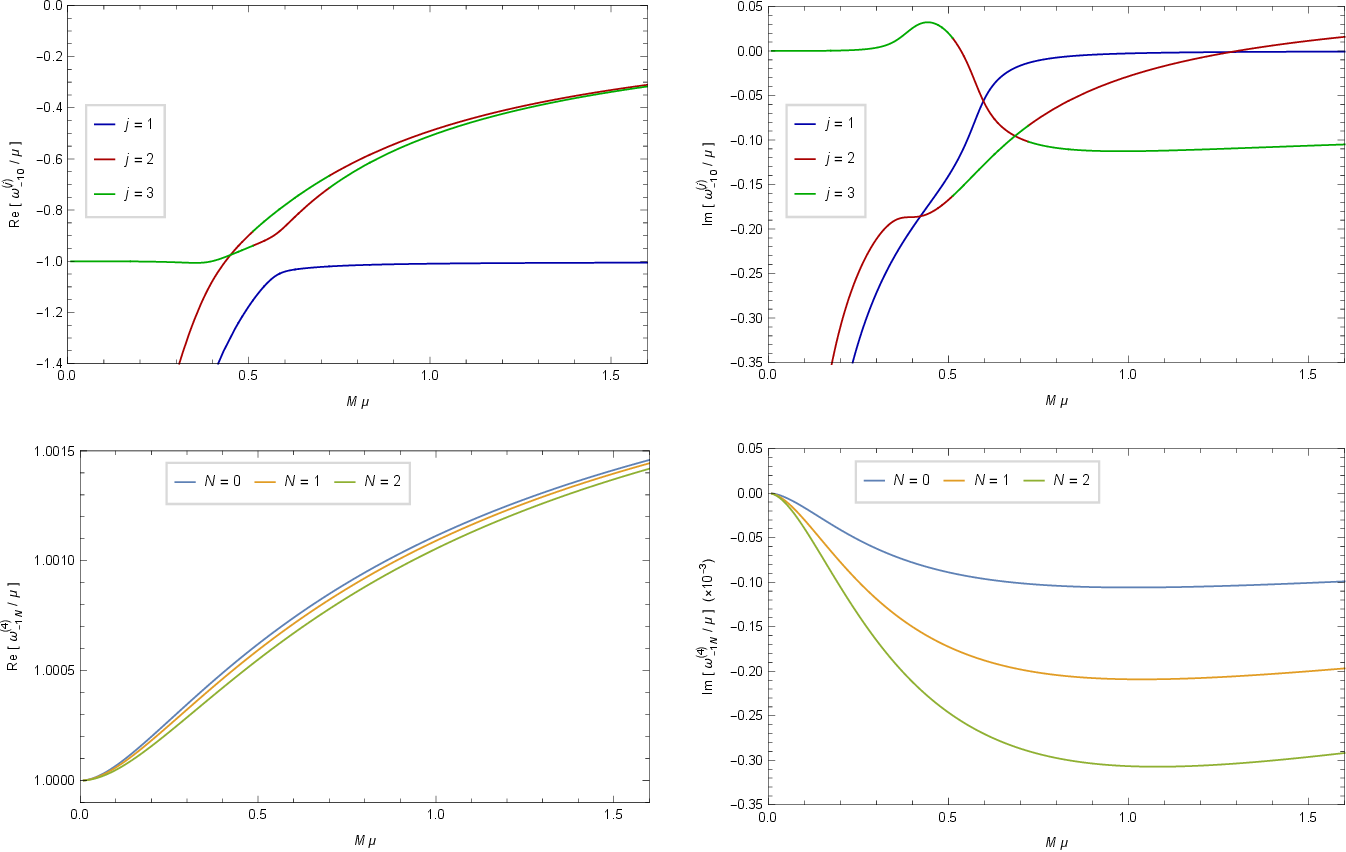}
\caption{Massive scalar resonant frequencies in the Kerr black hole spacetime for $m=-1$, and $a=0.99$. The left plots show the oscillation frequency $\mbox{Re}[\omega_{mN}^{(j)}/\mu]$, while the right plots show the decay (or growth) rate $\mbox{Im}[\omega_{mN}^{(j)}/\mu]$.}
\label{fig:Fig2}
\end{figure}

We also show the behavior of the massive scalar resonant frequencies $\omega_{mN}^{(j)}$ in Fig.~\ref{fig:Fig2}, as a function of the gravitational coupling $M\mu$. In Fig.~\ref{fig:Fig2} we see that the resonant frequencies $\omega_{mN}^{(4)}$ have a behavior which is similar to the results presented in Fig.~4 of Dolan's paper \cite{PhysRevD.76.084001}; here, in comparison with Dolan's paper, the overtone number $N$ ``plays the role'' of the parameter $m$. Anyway, it also indicates that the VBK approach generalizes the numerical approaches known in the literature (see Ref.~\cite{PhysRevD.76.084001} and references therein). In the present case, the spectrum has four frequency eigenvalues, which may describe the full behavior of massive scalar particles in the Kerr black hole spacetime. In addition, as in the Dolan's paper, ``zooming in'' on the upper right plot of Fig.~\ref{fig:Fig2} reveals the instability of the Kerr back hole, since the imaginary part of the resonant frequencies is actually positive at couplings $M\mu \lesssim 0.55$, and high rotation $a$ as well. Therefore, it is meaningful to study the instability of such a system for some interesting $m$ states, which is shown in Table \ref{tab:4_KNBH} and Fig.~\ref{fig:Fig4}, as functions of the gravitational coupling $M\mu$.

\begin{turnpage}
\begin{table}[ht]
\caption{Values of the resonant frequencies $\omega_{mN}^{(j)}$, and the real part of the corresponding coefficient $q_{mN}^{(j)}$, in the Kerr black hole spacetime for $m=0$, $a=0.1$, and $0.1 \leq M\mu \leq 1.0$. We focus on the fundamental mode $N=0$.}
\label{tab:2_KNBH}
\begin{tabular}{c|c|c|c|c|c|c|c|c}
\hline
\noalign{\smallskip}\hline\noalign{\smallskip}
$M\mu$ & $\omega_{00}^{(1)}$   & $\mbox{Re}[q_{00}^{(1)}]$ & $\omega_{00}^{(2)}$ & $\mbox{Re}[q_{00}^{(2)}]$ & $\omega_{00}^{(3)}$       & $\mbox{Re}[q_{00}^{(3)}]$ & $\omega_{00}^{(4)}$  & $\mbox{Re}[q_{00}^{(4)}]$ \\
\noalign{\smallskip}\hline\noalign{\smallskip}
0.1    & $-0.099567-0.000173i$ & $0.009465$                & $-0.249025i$        & $0.268353$                & $-99.24940i$              & $99.24940$                & $0.099567-0.000173i$ & $0.009465$                \\
0.2    & $-99.24940i$          & $99.24960$                & $-0.245380i$        & $0.316562$                & $-0.197536-0.001995i$     & $0.033494$                & $0.197536-0.001995i$ & $0.033494$                \\
0.3    & $-99.24940i$          & $99.24980$                & $-0.235450i$        & $0.381362$                & $-0.294439-0.006960i$     & $0.065768$                & $0.294439-0.006960i$ & $0.065768$                \\
0.4    & $-99.24940i$          & $99.25020$                & $-0.218579i$        & $0.455825$                & $-0.391276-0.015396i$     & $0.102823$                & $0.391276-0.015396i$ & $0.102823$                \\
0.5    & $-99.24940i$          & $99.25060$                & $-0.195165i$        & $0.536740$                & $-0.488716-0.027102i$     & $0.143053$                & $0.488716-0.027102i$ & $0.143053$                \\
0.6    & $-99.24940i$          & $99.25120$                & $-0.165829i$        & $0.622494$                & $-0.587135-0.041769i$     & $0.185653$                & $0.587135-0.041769i$ & $0.185653$                \\
0.7    & $-99.24940i$          & $99.25180$                & $-0.131168i$        & $0.712183$                & $-0.686736-0.059098i$     & $0.230168$                & $0.686736-0.059098i$ & $0.230168$                \\
0.8    & $-99.24940i$          & $99.25260$                & $-0.091697i$        & $0.805238$                & $-0.787628-0.078831i$     & $0.276316$                & $0.787628-0.078831i$ & $0.276316$                \\
0.9    & $-99.24940i$          & $99.25350$                & $-0.047852i$        & $0.901271$                & $-0.889862-0.100751i$     & $0.323901$                & $0.889862-0.100751i$ & $0.323901$                \\
1.0    & $-99.24940i$          & $99.25440$                & $\ 0.000000$        & $1.000000$                & $-0.993463-0.124673i$     & $0.372783$                & $0.993463-0.124673i$ & $0.372783$                \\
\noalign{\smallskip}\hline\noalign{\smallskip}
\hline
\end{tabular}
\end{table}
\end{turnpage}

%
%
\subsection{Kerr-Newman black hole}
Now, we present the spectrum of quasibound state frequencies in the Kerr-Newman black hole spacetime. In this case, the characteristic resonance equation (\ref{eq:characteristic_resonance_equation_KNBH}) has four complex solutions, which are denoted by $\omega_{mN}^{(j)}$, where $j=1,2,3,4$ labels the solutions.

Some values of the charged massive scalar resonant frequencies in the Kerr-Newman black hole, as well as the corresponding coefficient $q_{mN}^{(j)}$, are shown in Table \ref{tab:3_KNBH}, as functions of the gravitational coupling $M\mu$. From Table \ref{tab:3_KNBH}, we conclude that all resonant frequencies $\omega_{mN}^{(j)}$ are physically admissible, since $\mbox{Re}[q_{mN}^{(j)}] > 0$ and hence they represent the spectrum of quasibound state frequencies for charged massive scalar particles in the Kerr-Newman black hole spacetime. However, as Eq.~(\ref{eq:resonance_equation_KNBH}) seems to have just two solutions, we may conclude that the first $\omega_{mN}^{(1)}$ and second solutions $\omega_{mN}^{(2)}$ are incidental in this scenario. In addition, all solutions are different from each other, in this scenario.

\begin{figure}[p]
\centering
\includegraphics[width=1\columnwidth]{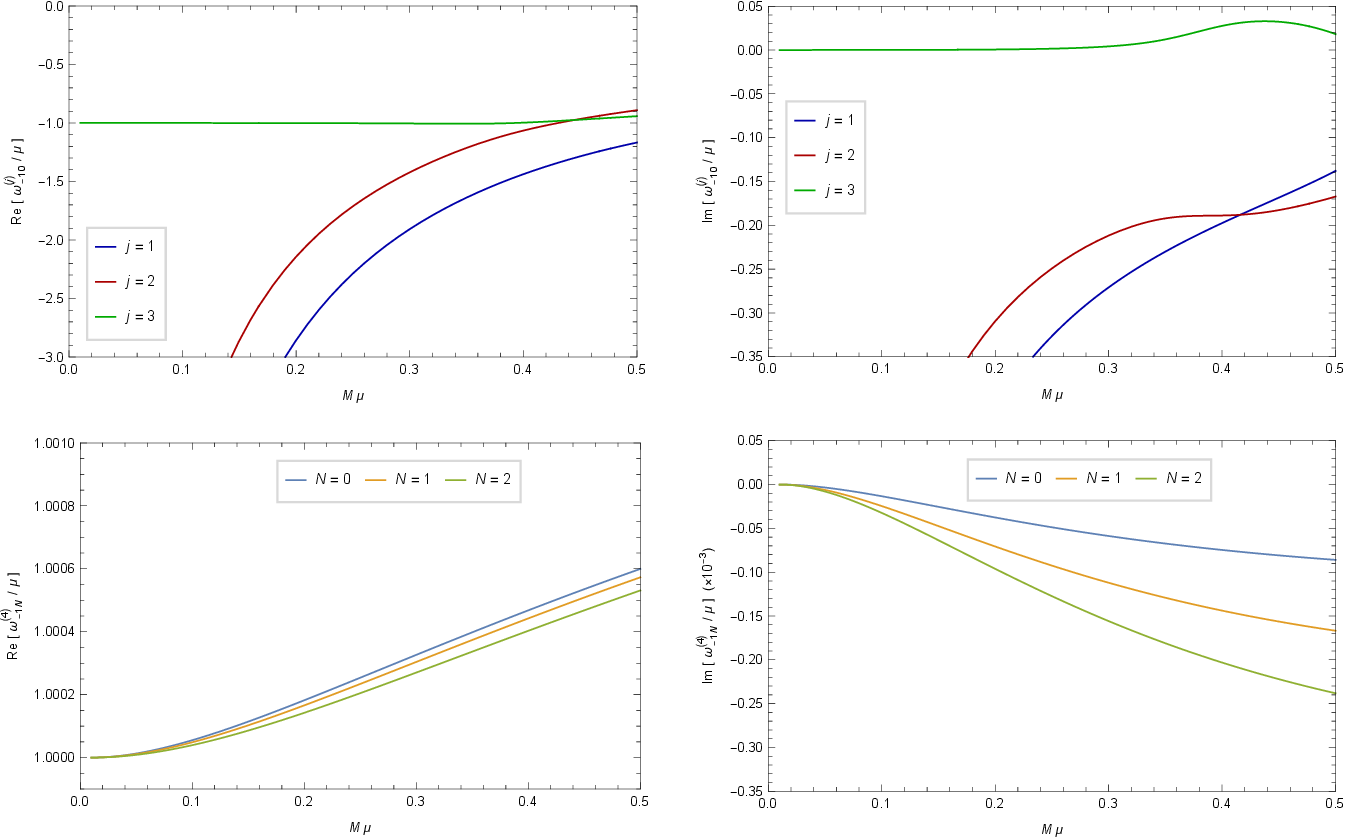}
\caption{Massive scalar resonant frequencies in the Kerr-Newman black hole spacetime for $m=-1$, $a=0.99$, $e=1$, and $Q=0.01$. The left plots show the oscillation frequency $\mbox{Re}[\omega_{mN}^{(j)}/\mu]$, while the right plots show the decay (or growth) rate $\mbox{Im}[\omega_{mN}^{(j)}/\mu]$.}
\label{fig:Fig3}
\end{figure}

In order to compare our analytical results with the numerical ones obtained by Furuhashi and Nambu \cite{ProgTheorPhys.112.983}, we show the behavior of the charged massive scalar resonant frequencies $\omega_{mN}^{(j)}$ in Fig.~\ref{fig:Fig3}, as a function of the gravitational coupling $M\mu$. In Fig.~\ref{fig:Fig3} we see that the resonant frequencies $\omega_{mN}^{(3)}$ have a behavior which is similar to the results presented in Fig.~6 of Furuhashi and Nambu's paper \cite{ProgTheorPhys.112.983}. Once again, it indicates that the VBK approach generalizes the numerical approaches known in the literature (see Ref.~\cite{ProgTheorPhys.112.983} and references therein).

\begin{figure}[p]
\centering
\includegraphics[width=1\columnwidth]{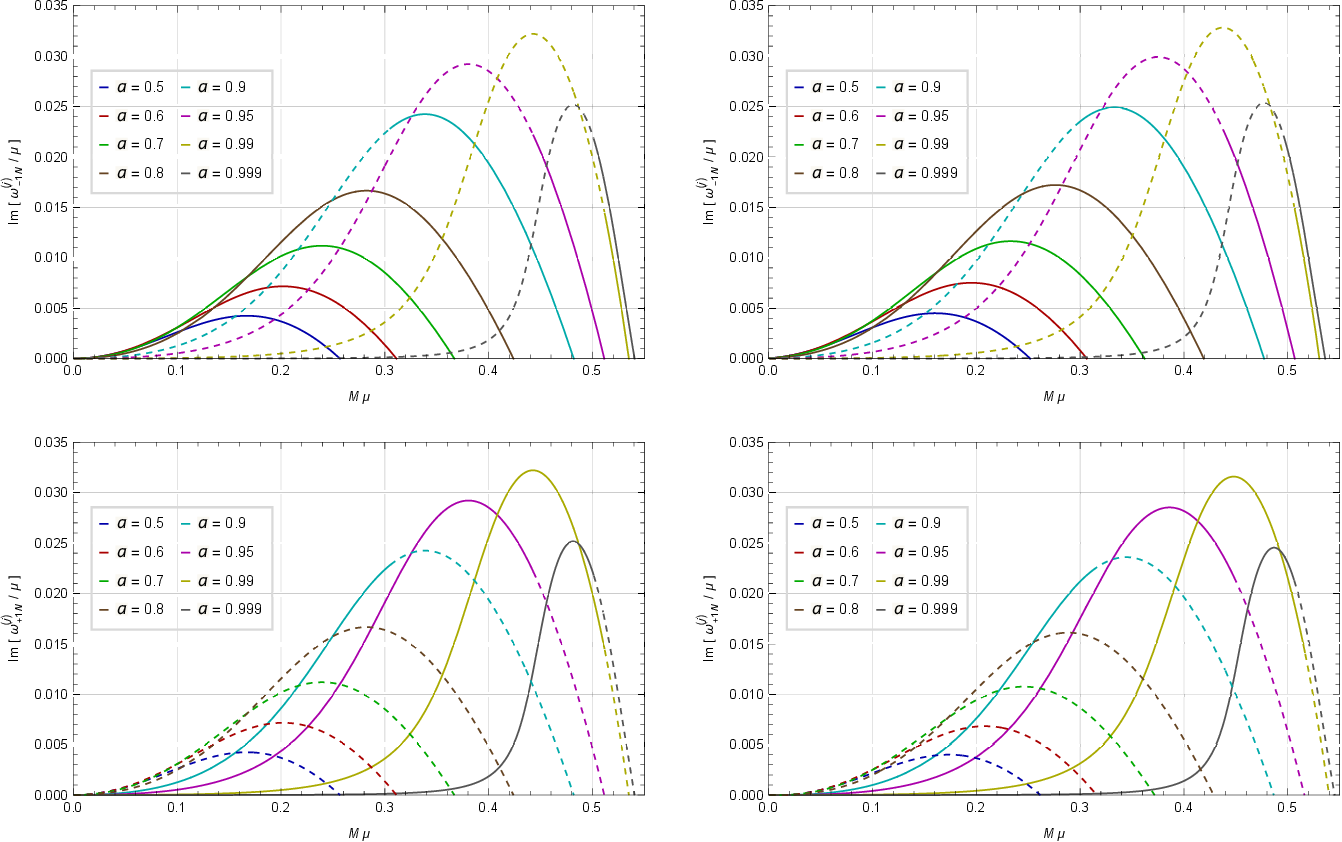}
\caption{Instability of the $m=-1,+1$ states for $N=0$. The growth rate of the quasibound states is shown as a function of the gravitational coupling $M\mu$, for a range of black hole rotations $a$. The left plots show the instability of a Kerr black hole, while the right plots are related to the Kerr-Newman black hole for $e=1$ and $Q=0.01$. The solid and dashed lines represent the resonant frequencies $\omega_{mN}^{(2)}$ and $\omega_{mN}^{(3)}$, respectively.}
\label{fig:Fig4}
\end{figure}

In addition, as in the Furuhashi and Nambu's paper, ``zooming in'' on the upper right plot of Fig.~\ref{fig:Fig3} reveals the instability of the Kerr-Newman back hole, since the imaginary part of the resonant frequencies is actually positive at couplings $M\mu \lesssim 0.5$, and high rotation $a$ as well. The instability of such a system for some interesting $m$ states is presented in Table \ref{tab:4_KNBH} and Fig.~\ref{fig:Fig4}, as functions of the gravitational coupling $M\mu$. From Table \ref{tab:4_KNBH} and Fig.~\ref{fig:Fig4}, we can conclude that the maximum instability growth rates of the Kerr black hole do not depend on the sign of the magnetic quantum number $m$, that is, they are the same for $m=-1$ and $m=+1$. On the other hand, the maximum instability growth rates of the Kerr-Newman black hole are sensitive to the sign of $m$, since their peaks are highest for $m=-1$.

It is worth emphasizing that these results were obtained directly from the exact, analytic and general solution of the charged massive Klein-Gordon equation in the spacetime under consideration, which is given in terms of the confluent Heun functions. Therefore, we did not use any numerical approximations to obtain these values and graphs, as it is commonly used in the literature. Furthermore, our results are valid in the entire range $M\mu \geq 0$.

\begin{turnpage}
\begin{table}[ht]
\caption{Values of the resonant frequencies $\omega_{mN}^{(j)}$, and the real part of the corresponding coefficient $q_{mN}^{(j)}$, in the Kerr-Newman black hole spacetime for $m=0$, $a=0.1$, $e=0.01$, $Q=0.1$, and $0.1 \leq M\mu \leq 1.0$. We focus on the fundamental mode $N=0$.}
\label{tab:3_KNBH}
\begin{tabular}{c|c|c|c|c|c|c|c|c}
\hline
\noalign{\smallskip}\hline\noalign{\smallskip}
$M\mu$ & $\omega_{00}^{(1)}$   & $\mbox{Re}[q_{00}^{(1)}]$ & $\omega_{00}^{(2)}$  & $\mbox{Re}[q_{00}^{(2)}]$ & $\omega_{00}^{(3)}$       & $\mbox{Re}[q_{00}^{(3)}]$ & $\omega_{00}^{(4)}$  & $\mbox{Re}[q_{00}^{(4)}]$ \\
\noalign{\smallskip}\hline\noalign{\smallskip}
0.1    & $-0.099559-0.000177i$ & $0.009554$                & $0.000995-98.00500i$ & $98.00500$                & $0.000485-0.249017i$       & $0.268345$                & $0.099575-0.000169i$ & $0.009375$                \\
0.2    & $-0.197517-0.002016i$ & $0.033637$                & $0.197556-0.001974i$ & $0.033350$                & $0.000461-0.245372i$       & $0.316556$                & $0.000995-98.00500i$ & $98.00520$                \\
0.3    & $-0.294413-0.007005i$ & $0.065941$                & $0.294465-0.006916i$ & $0.065592$                & $0.000448-0.235442i$       & $0.381356$                & $0.000995-98.00500i$ & $98.00540$                \\
0.4    & $-0.391250-0.015464i$ & $0.103016$                & $0.391304-0.015328i$ & $0.102625$                & $0.000446-0.218571i$       & $0.455821$                & $0.000995-98.00500i$ & $98.00580$                \\
0.5    & $-0.488694-0.027194i$ & $0.143260$                & $0.488744-0.027011i$ & $0.142835$                & $0.000449-0.195157i$       & $0.536736$                & $0.000996-98.00500i$ & $98.00630$                \\
0.6    & $-0.587118-0.041882i$ & $0.185872$                & $0.587160-0.041656i$ & $0.185417$                & $0.000457-0.165822i$       & $0.622492$                & $0.000996-98.00500i$ & $98.00680$                \\
0.7    & $-0.686727-0.059231i$ & $0.230397$                & $0.686759-0.058964i$ & $0.229915$                & $0.000467-0.131162i$       & $0.712182$                & $0.000997-98.00500i$ & $98.00750$                \\
0.8    & $-0.787627-0.078982i$ & $0.276554$                & $0.787647-0.078677i$ & $0.276044$                & $0.000478-0.091692i$       & $0.805237$                & $0.000998-98.00500i$ & $98.00830$                \\
0.9    & $-0.889871-0.100919i$ & $0.324146$                & $0.889878-0.100578i$ & $0.323611$                & $0.000490-0.047850i$       & $0.901271$                & $0.000999-98.00500i$ & $98.00910$                \\
1.0    & $-0.993483-0.124857i$ & $0.373034$                & $0.993476-0.124481i$ & $0.372474$                & $0.000502+6.1553\mbox{e-8}$     & $1.000000$                & $0.001000-98.00500i$ & $98.01010$                \\
\noalign{\smallskip}\hline\noalign{\smallskip}
\hline
\end{tabular}
\end{table}
\end{turnpage}

\begin{turnpage}
\begin{table}[ht]
\caption{Maximum instability growth rates, $\tau^{-1} \sim \mbox{Im}[\omega_{mN}^{(j)}/\mu]$, in the Kerr and Kerr-Newman (with $e=1$ and $Q=0.01$) black holes. We focus on the $N=0$ and $m=-1,+1$ states.}
\label{tab:4_KNBH}
\begin{tabular}{c|c|c|c|c|c|c|c|c}
\hline
\noalign{\smallskip}\hline\noalign{\smallskip}
        & \multicolumn{4}{c}{Kerr}\vline                                      & \multicolumn{4}{c}{Kerr-Newman}                        \\
        & \multicolumn{2}{c}{$m=-1$}\vline & \multicolumn{2}{c}{$m=+1$}\vline & \multicolumn{2}{c}{$m=-1$}\vline & \multicolumn{2}{c}{$m=+1$} \\
$a$     & $M\mu$     & $\tau^{-1}$         & $M\mu$   & $\tau^{-1}$           & $M\mu$     & $\tau^{-1}$         & $M\mu$     & $\tau^{-1}$     \\
\noalign{\smallskip}\hline\noalign{\smallskip}
$0.5$   & $0.166885$ & $0.004246$          & $0.166885$ & $0.004246$          & $0.160316$ & $0.004500$          & $0.173477$ & $0.004003$    \\
$0.6$   & $0.201837$ & $0.007167$          & $0.201837$ & $0.007167$          & $0.195359$ & $0.007517$          & $0.208343$ & $0.006830$    \\
$0.7$   & $0.239344$ & $0.011192$          & $0.239344$ & $0.011192$          & $0.233015$ & $0.011647$          & $0.245710$ & $0.010753$    \\
$0.8$   & $0.282288$ & $0.016656$          & $0.282288$ & $0.016656$          & $0.276186$ & $0.017221$          & $0.288437$ & $0.016109$    \\
$0.9$   & $0.338554$ & $0.024248$          & $0.338554$ & $0.024248$          & $0.332809$ & $0.024919$          & $0.344368$ & $0.023597$    \\
$0.95$  & $0.380566$ & $0.029213$          & $0.380566$ & $0.029213$          & $0.375095$ & $0.029917$          & $0.386141$ & $0.028530$    \\
$0.99$  & $0.442628$ & $0.032213$          & $0.442628$ & $0.032213$          & $0.437558$ & $0.032835$          & $0.447966$ & $0.031589$    \\
$0.999$ & $0.481321$ & $0.025162$          & $0.481321$ & $0.025162$          & $0.476740$ & $0.025359$          & $0.486849$ & $0.024539$    \\
\noalign{\smallskip}\hline\noalign{\smallskip}
\hline
\end{tabular}
\end{table}
\end{turnpage}

%
%
\section{Wave eigenfunctions}\label{Wave_functions}
In this section we will derive both the angular and radial wave eigenfunctions related to the quasibound states of charged massive scalar particles propagating in the Kerr-Newman black hole spacetime, as well as in the particular cases of Schwarzschild and Kerr backgrounds.
%
%
\subsection{Radial eigenfunctions}\label{Radial_eigenfunctions}
We will use some properties of the confluent Heun functions and then obtains their polynomial expressions which describes the radial wave eigenfunctions related to the scalar quasibound states in the background under consideration.

To do this, it is worth noticing that the two polynomial conditions, given by Eqs.~(\ref{eq:delta-condition}) and (\ref{eq:Delta-condition}), are consistent if and only if the parameter $\xi$, given by Eq.~(\ref{eq:sigma0_confluent_Heun}), is chosen properly, which means that is calculated via a polynomial equation of degree $N+1$, namely, $c_{N+1}=0$, where the coefficients $c_{n}$ are given by Eq.~(\ref{eq:cn}). Thus, since we will have $s(=1,2,\ldots,N+1)$ solutions for $\xi$, we can use the notation $\xi_{mN;s}$ for these eigenvalues. In our case, the corresponding confluent Heun polynomials will be denoted as $\mbox{HeunCp}_{mN;s}(x)$. As a consequence, we will fix the values of the separation constant $\lambda$ according to each value of the overtone number $N$ and resonant frequency $\omega_{mN}^{(j)}$, so that we will use the notation $\lambda_{mN;s}^{(j)}$ for these eigenvalues.

Let us obtain the explicit form of the first two confluent Heun polynomials, as follows. For $N=0$, we have
\begin{equation}
\mbox{HeunCp}_{m0;s}(x)=c_{0}=1.
\label{eq:HCp0N}
\end{equation}
The eigenvalues $\xi_{m0;s}$ must obey
\begin{equation}
c_{1}=0,
\end{equation}
which implies
\begin{equation}
c_{1}=\frac{T_{1}}{P_{1}},
\label{eq:c1}
\end{equation}
and then we get
\begin{equation}
\xi_{m0;1}=0.
\label{eq:sigma000}
\end{equation}
Thus, the confluent Heun polynomial for the fundamental mode is given by
\begin{equation}
\mbox{HeunCp}_{m0;1}(x)=1.
\label{eq:HCp00}
\end{equation}

Next, for $N=1$, we have
\begin{equation}
\mbox{HeunCp}_{m1;s}(x)=c_{0}+c_{1}x=1-\frac{\xi_{m1;s}}{1+\beta}x.
\label{eq:Hp1m}
\end{equation}
The eigenvalues $\xi_{m1;s}$ must obey
\begin{equation}
c_{2}=0,
\end{equation}
where
\begin{equation}
P_{2}c_{2}=T_{2}c_{1}+X_{2}c_{0},
\label{eq:P1}
\end{equation}
which implies that
\begin{equation}
\xi_{m1;s}=\frac{-(\alpha-2-\beta-\gamma) \pm \sqrt{\Delta}}{2},
\label{eq:sigma01N}
\end{equation}
with $\Delta=(\alpha-2-\beta-\gamma)^{2}+4\alpha(1+\beta)$. Here, the signs $-,+$ stand for $s=1,2$, respectively, that is, the first and second solution for $\xi$. Thus, the confluent Heun polynomials for the first excited mode are given by
\begin{equation}
\mbox{HeunCp}_{m1;1}(x)=1-\frac{-(\alpha-2-\beta-\gamma) - \sqrt{\Delta}}{2(1+\beta)}x,
\label{eq:HCp10}
\end{equation}
\begin{equation}
\mbox{HeunCp}_{m1;2}(x)=1-\frac{-(\alpha-2-\beta-\gamma) + \sqrt{\Delta}}{2(1+\beta)}x.
\label{eq:HCp11}
\end{equation}
The same procedure can be used to fix the values of the separation constant $\lambda$ for $N=1,2,\ldots$. However, there are too many final values that no insight is gained by writing them out and hence we focus on the fundamental mode.

Therefore, the radial wave eigenfunctions, related to the quasibound states of charged massive scalar particles propagating in a Kerr-Newman black hole spacetime, can be written as
\begin{equation}
^{(j)}R_{mN;s}(x)=C_{mN;s}^{(j)}\ x^{D_{0}}(x-1)^{D_{1}}\mbox{e}^{D_{2}x}\ \mbox{HeunCp}_{mN;s}^{(j)}(x),
\label{eq:eigenfunctions_KNBH}
\end{equation}
where $j=1,2,3,4$ labels the resonant frequency solutions in this background, and $C_{mN;s}^{(j)}$ is a constant to be determined by using some additional boundary condition, as for example, that the wave function should be appropriately normalized in the range between the exterior event horizon and infinity. This expression is also valid to massive scalar particles in a Kerr spacetime.

On the other hand, in the case of a Schwarzschild black hole spacetime, the radial wave eigenfunctions are given by
\begin{equation}
^{(j)}R_{N;s}(x)=C_{N;s}^{(j)}\ x^{D_{0}}(x-1)^{D_{1}}\mbox{e}^{D_{2}x}\ \mbox{HeunCp}_{N;s}^{(j)}(x),
\label{eq:eigenfunctions_SBH}
\end{equation}
where $j=1,2,3$ labels the resonant frequency solutions in this background, and $C_{N;s}^{(j)}$ is a constant to be determined.

\begin{figure}[p]
\centering
\includegraphics[width=1\columnwidth]{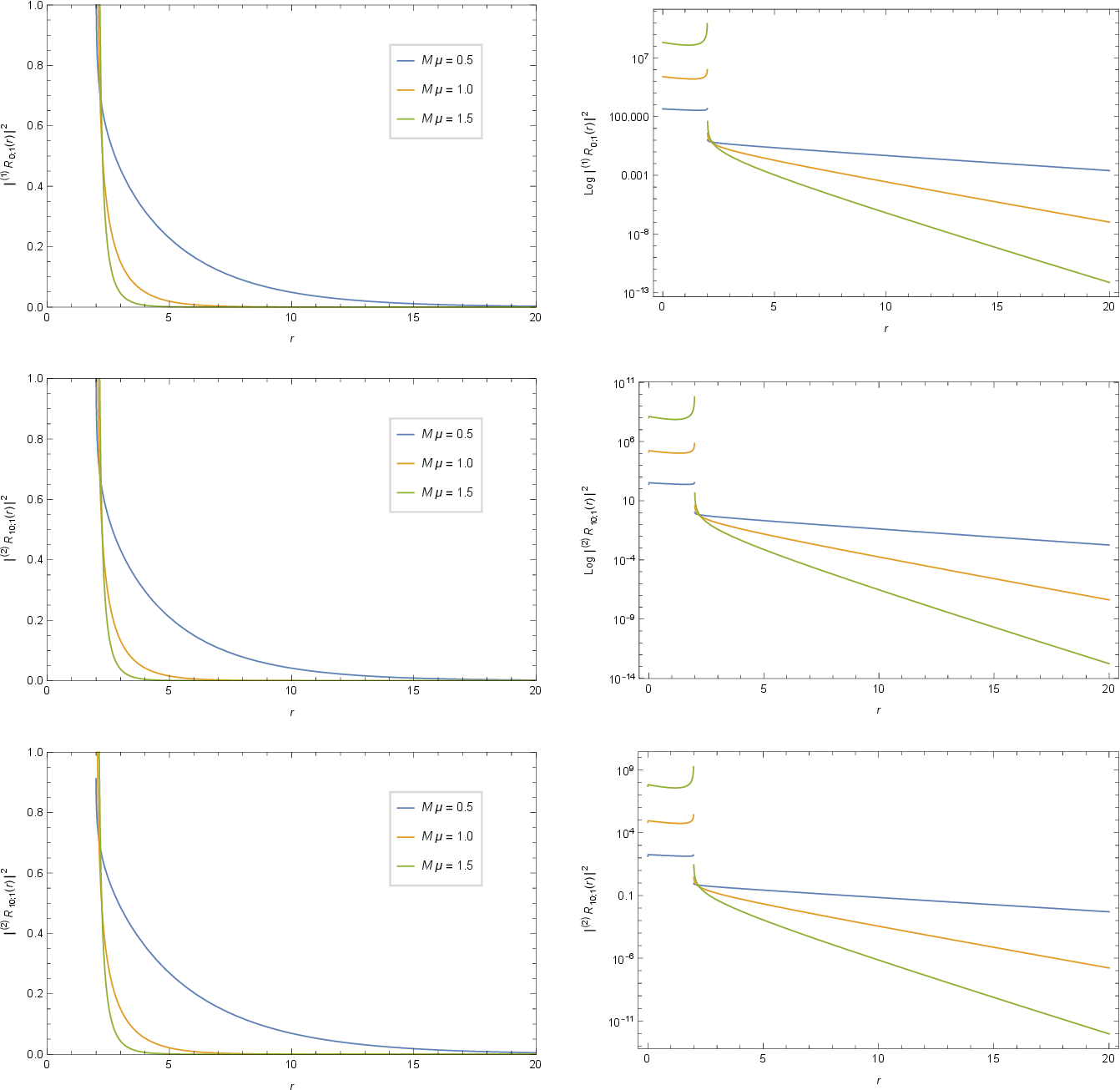}
\caption{The squared radial wave eigenfunctions for the fundamental mode $N=0$. On the top, we show it in the Schwarzschild black hole. On the middle, we present it in the Kerr black hole for $m=1$, and $a=0.1$. On the bottom, we get it in the Kerr-Newman black hole for $m=1$, $a=0.1$, $e=1$, and $Q=0.1$. The units are in multiples of $C_{mN;s}^{(j)}$.}
\label{fig:Fig5}
\end{figure}

Thus, by using Eqs.~(\ref{eq:eigenfunctions_KNBH}) and (\ref{eq:eigenfunctions_SBH}), we can plot the squared radial wave eigenfunctions for the fundamental mode, which are presented in Fig.~\ref{fig:Fig5}. From Fig.~\ref{fig:Fig5}, we conclude that the radial wave eigenfunction for the fundamental mode $R_{0;1}^{(1)}$ describes quasibound states of massive scalar particles in the Schwarzschild black hole, as well as $R_{10;1}^{(2)}$ in the Kerr and Kerr-Newman black holes, since they present the desired behavior, namely, the radial solution tends to zero at infinity and diverges at the exterior event horizon; it mathematically reaches a maximum value and then crosses into the black hole.

%
%
\subsection{Angular eigenfunctions}\label{Angular_eigenfunctions}
Now, it is known that the regularity requirements of the angular functions at the two boundaries $\theta=0$ and $\theta=\pi$ single out a discrete set of angular eigenvalues $\lambda$, which couples the angular equation (\ref{eq:angular_KNBH}) and the radial equation (\ref{eq:radial_KNBH}). In the present case, we can obtain an expression for the eigenvalues $\lambda$ from the $\Delta$-condition \cite{PhysRevD.105.045015,EurPhysJC.82.669} given by Eq.~(\ref{eq:Delta-condition}), which is equivalent to $c_{N+1}=0$ in Eq.~(\ref{eq:cn}).

Thus, for the fundamental mode $N=0$, the parameter $\xi$ must obey the relation $c_{1}=0$, and then we have that $\xi=0$. Therefore, from Eq.~(\ref{eq:sigma0_confluent_Heun}), we obtain the following expression for the eigenvalues $\lambda_{mN}^{(j)}$:
\begin{eqnarray}
\lambda_{m0}^{(j)}	& = & \frac{2 a m Q e -m^2 (Q^2-1)+Q^2 e ^2}{a^2+Q^2-1}+\frac{i (a m + Q e)}{\sqrt{1-a^2-Q^2}}+2 \mu ^2 (\sqrt{1-a^2-Q^2}-1)+m^2+\mu ^2 Q^2+Q^2 e ^2\nonumber\\
										&		& -\frac{[2 Q e  (1-a^2-Q^2)^{3/2}+i (Q^2 -2) \sqrt{1-a^2-Q^2}+6 a^2 Q e -2 a m Q^2+4 a m+4 Q^3 e -2 Q e ]\omega}{a^2+Q^2-1}\nonumber\\
										&		& +\biggl[\frac{(Q^2-2)^2}{a^2+Q^2-1}-4 \sqrt{1-a^2-Q^2}-2 Q^2+8\biggr]\omega^2\nonumber\\
										&		& +2 \sqrt{\mu ^2-\omega ^2} [(2 i \omega - i Q e + 1) \sqrt{1-a^2-Q^2}+i a m+i Q^2 \omega +i Q e -2 i \omega],
\label{eq:lambda_KNBH}
\end{eqnarray}
where $\omega=\omega_{m0}^{(j)}$. As it was expected, the eigenvalues $\lambda_{m0}^{(j)}$ are complex; the application of complex angular momentum (CAM) techniques to atomic and molecular scattering by evaluating the Legendre functions of complex degree $\nu$ was reported in the end of 1970's \cite{Connor:1979MP}, as well as in the context of black hole scattering in the 1990's \cite{ClassQuantumGrav.11.2991,ClassQuantumGrav.11.3003}. Thus, by using the Wolfram Mathematica$^{\mbox{\textregistered}}$ version 12.3, the complex degree $\nu$ can be numerically evaluated for both associated Legendre and spheroidal functions.

It is worth emphasizing that the two numerically satisfactory solutions of the spheroidal and associated Legendre equations of complex degree are given in terms of the Ferrers function of the first kind $Ps_{\nu}^{-m}(-z)$ and $Ps_{\nu}^{-m}(z)$, and $P_{\nu}^{-m}(-z)$ and $P_{\nu}^{-m}(z)$, respectively, in the interval $-1 \leq z \leq 1$.

\begin{figure}[p]
\centering
\includegraphics[width=0.45\columnwidth]{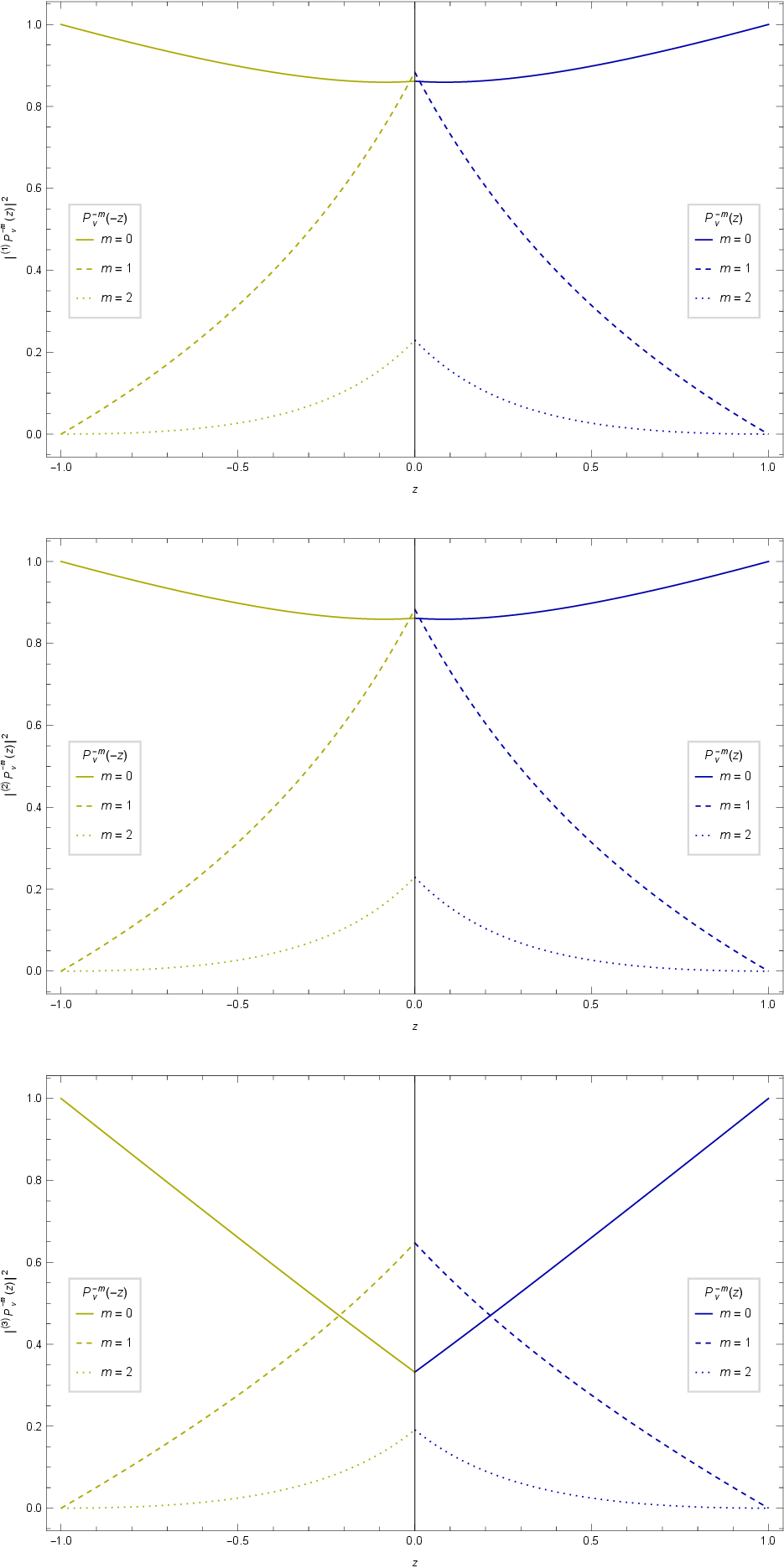}
\caption{The squared angular wave eigenfunctions in the Schwarzschild black hole spacetime for the fundamental mode $N=0$, where $z=\cos\theta$. The units are in multiples of $^{(j)}C_{\nu}^{-m}$.}
\label{fig:Fig6}
\end{figure}

\begin{figure}[p]
\centering
\includegraphics[width=0.45\columnwidth]{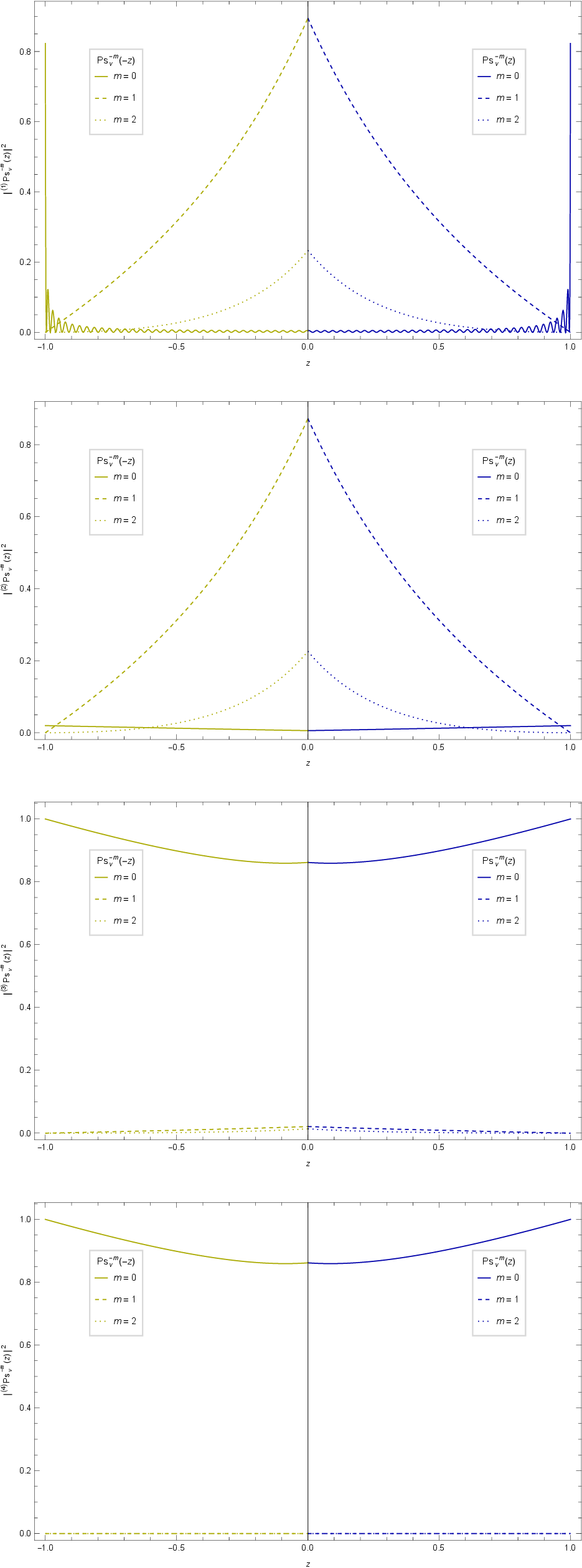}
\caption{The squared angular wave eigenfunctions in the Kerr black hole spacetime for the fundamental mode $N=0$, with $M\mu=0.5$, and $a=0.01$, where $z=\cos\theta$. The units are in multiples of $^{(j)}C_{\nu}^{-m}$.}
\label{fig:Fig7}
\end{figure}

\begin{figure}[p]
\centering
\includegraphics[width=0.45\columnwidth]{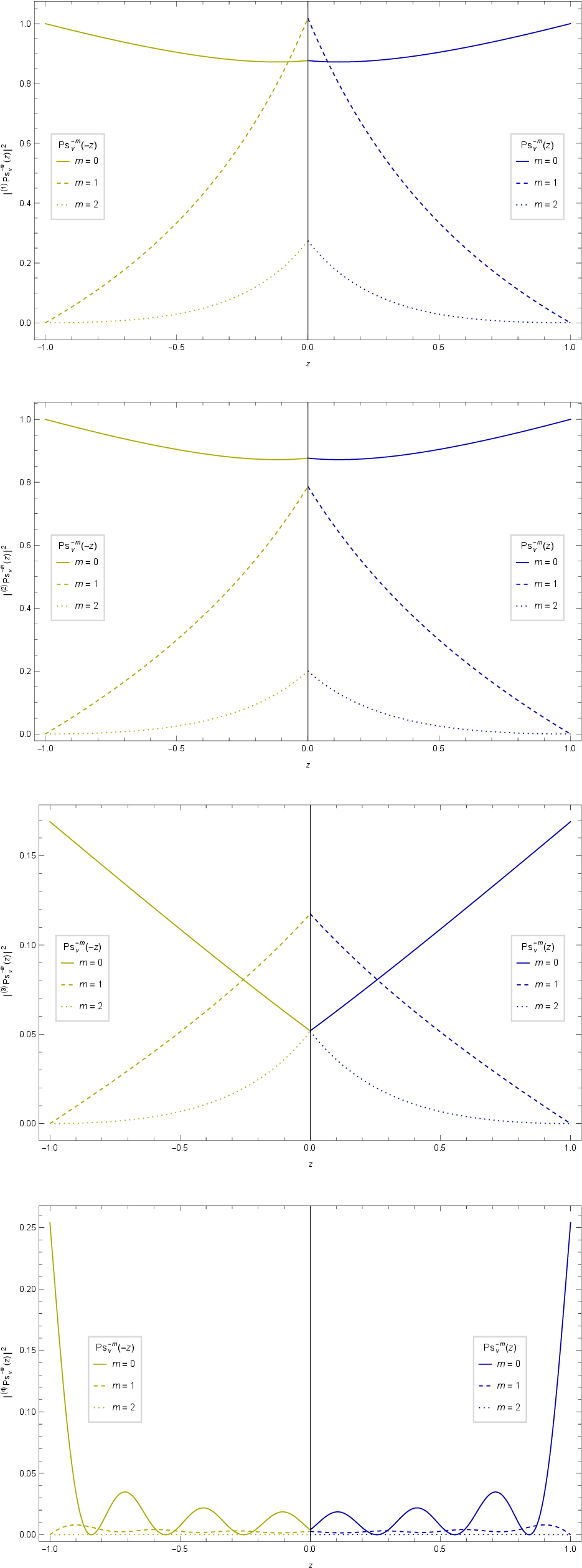}
\caption{The squared angular wave eigenfunctions in the Kerr-Newman black hole spacetime for the fundamental mode $N=0$, with $M\mu=0.5$, $a=0.1$, $e=0.01$, and $Q=0.1$, where $z=\cos\theta$. The units are in multiples of $^{(j)}C_{\nu}^{-m}$.}
\label{fig:Fig8}
\end{figure}

In Tables \ref{tab:5_KNBH}-\ref{tab:7_KNBH}, we present the some eigenvalues $\lambda_{m0}^{(j)}$, as well as the corresponding complex degree $\nu_{m0}^{(j)}$, as a function of the azimuthal quantum number $m$, for fixed values of the gravitational coupling $M\mu$ and angular momentum $a$. In addition, we also present the behavior of the angular eigenfunctions $Ps_{\nu}^{-m}(z=\cos\theta)$ in Figs.~\ref{fig:Fig6}-\ref{fig:Fig8}, for some values of the azimuthal quantum number $m$. From Figs.~\ref{fig:Fig6}-\ref{fig:Fig8}, we conclude that the angular wave eigenfunctions for the fundamental mode $N=0$ are regular (finite) at $z=\pm 1$ and therefore it describes scalar quasibound states in the background under consideration.

\begin{turnpage}
\begin{table}[ht]
\caption{The eigenvalue $\lambda_{0}^{(j)}$, and the corresponding complex degree $\nu_{0}^{(j)}$, in the Schwarzschild black hole spacetime.}
\label{tab:5_KNBH}
\begin{tabular}{c||c|c||c|c||c|c}
\hline
\noalign{\smallskip}\hline\noalign{\smallskip}
	$M\mu$ & $\lambda_{0}^{(1)}$  & $\nu_{0}^{(1)}$      & $\lambda_{0}^{(2)}$  & $\nu_{0}^{(2)}$      & $\lambda_{0}^{(3)}$ & $\nu_{0}^{(3)}$ \\
	\noalign{\smallskip}\hline\noalign{\smallskip}
	0      & $0$					        &	$0$                  & $0$                  & $0$                  & $0$                 & $0$        \\
	0.5    & $0.232786-0.792552i$	&	$0.339883-0.471823i$ & $0.232786+0.792552i$ & $0.339883+0.471823i$ & $0.682328$          & $0.465571$	\\
	1.0    & $0.500000-1.322880i$	&	$0.565526-0.620762i$ & $0.500000+1.322880i$ & $0.565526+0.620762i$ & $2.000000$          & $1.000000$	\\
\noalign{\smallskip}\hline\noalign{\smallskip}
\hline
\end{tabular}
\end{table}
\end{turnpage}

\begin{turnpage}
\squeezetable
\begin{table}[ht]
\caption{The eigenvalue $\lambda_{m0}^{(j)}$, and the corresponding complex degree $\nu_{m0}^{(j)}$, in the Kerr black hole spacetime, for $m=1$, and $a=0.99$.}
\label{tab:6_KNBH}
\begin{tabular}{c||c|c||c|c||c|c||c|c}
\hline
\noalign{\smallskip}\hline\noalign{\smallskip}
	$M\mu$ & $\lambda_{10}^{(1)}$  & $\nu_{10}^{(1)}$     & $\lambda_{10}^{(2)}$  & $\nu_{10}^{(2)}$     & $\lambda_{10}^{(3)}$            & $\nu_{10}^{(3)}$     & $\lambda_{10}^{(4)}$  & $\nu_{10}^{(4)}$      \\
	\noalign{\smallskip}\hline\noalign{\smallskip}
	0      & $-49.25130+7.017920i$ & $7.017920i$          & $-49.25130+7.017920i$ & $7.017920i$          & $0.301739-0.991344i$            & $0.395014-0.572708i$ & $1.282130+0.827407i$  & $0.735149+0.401333i$  \\
	0.5    & $-197.6270+14.09340i$ & $0.001255+14.05810i$ & $1.359450+0.520220i$  & $0.791228+0.195441i$ & $0.760894-1.328170i$            & $0.650458-0.554208i$ & $0.378751+0.701396i$  & $0.401302+0.422672i$  \\
	1.0    & $-443.9310+21.14810i$ & $0.001855+21.06970i$ & $0.923354-0.184643i$  & $0.863615-0.021191i$ & $-0.017036+0.059929i$           & $0.668590+0.014134i$ & $-47.17610-6.757170i$ & $-0.008302-6.868560i$ \\
\noalign{\smallskip}\hline\noalign{\smallskip}
\hline
\end{tabular}
\end{table}
\end{turnpage}

\begin{turnpage}
\squeezetable
\begin{table}[ht]
\caption{The eigenvalue $\lambda_{m0}^{(j)}$, and the corresponding complex degree $\nu_{m0}^{(j)}$, in the Kerr-Newman black hole spacetime, for $m=0$, $a=0.1$, $e=0.01$, and $Q=0.1$.}
\label{tab:7_KNBH}
\begin{tabular}{c||c|c||c|c||c|c||c|c}
\hline
\noalign{\smallskip}\hline\noalign{\smallskip}
	$M\mu$ & $\lambda_{00}^{(1)}$  & $\nu_{00}^{(1)}$             & $\lambda_{00}^{(2)}$  & $\nu_{00}^{(2)}$     & $\lambda_{00}^{(3)}$              & $\nu_{00}^{(3)}$               & $\lambda_{00}^{(4)}$  & $\nu_{00}^{(4)}$ \\
	\noalign{\smallskip}\hline\noalign{\smallskip}
	0      & $0.001010i$           & $1.0204\mbox{e-6}+0.001010i$ & $-0.007524+0.000989i$ & $0.499949$           & $0.001010i$              & $1.0204\mbox{e-6}+0.001010i$   & $1.970000-0.000979i$  & $6.571060$       \\
	0.5    & $0.222336+0.800634i$  & $0.337240+0.478056i$         & $0.221895-0.799381i$  & $0.336719-0.477605i$ & $0.666527+0.000430i$              & $0.497670-0.000015i$           & $1.967470-0.000979i$  & $6.570970$       \\
	1.0    & $0.463055+1.342930i$  & $0.556642+0.634875i$         & $0.462764-1.341900i$  & $0.556324-0.634578i$ & $1.969800+9.9491\mbox{e-8}i$          & $0.991233 + 3.3386\mbox{e-6}i$ & $1.959900-0.000979i$  & $6.570690$       \\
\noalign{\smallskip}\hline\noalign{\smallskip}
\hline
\end{tabular}
\end{table}
\end{turnpage}

%
%
\section{The supermassive black hole at the center of M87}\label{supermassive}
Lastly, we will apply our previous results for an actual astrophysical object, by computing the decay time of a ultra-light bosonic particle around the supermassive Kerr black hole situated at the center of M87 galaxy, whose shadow was recently observed \cite{ETH1}. We will consider the relativistic approximation $\omega/\mu \gg 1$. In this case, according to Eq.~(\ref{eq:characteristic_resonance_equation_KNBH}), we have
\begin{equation}\label{Rel-Kerr}
\omega_I\approx -\frac{c^3 (n+1)\sqrt{1-(a^{*})^2}} {GM(a^{*})^2},
\end{equation}
which does not depend on the particle mass, $\mu$, nor on the angular quantum number, $m$. It is worth calling attention to the fact that in Eq.~(\ref{Rel-Kerr}), we have reintroduced the fundamental constants and taken into account the fact that $M\mu\ll 1$, since $M\approx5 \times 10^9$ solar masses, $\mu\approx 10^{-21}$ eV/$c^2$, and the dimensionless parameter $a^{*}=a/MG\approx 0.9$ \cite{Denton}.

Therefore, the decay characteristic time for these ultra light bosons is given by
\begin{equation}\label{Time-Non-Rel}
\tau \approx \frac{1.3 \times 10^{39}}{(n+1)} \textrm{ seconds}.
\end{equation}
The huge magnitude of this time represents a stability for the system, even for large principal quantum numbers.
%
%
\section{Final remarks}\label{Final_remarks}
In this paper we obtained general analytical solutions for the charged massive Klein-Gordon equation in a Kerr-Newman black hole spacetime, as compared to the ones obtained in our previous work \cite{AnnPhys.350.14}, thus revisiting these later ones. In addition, we get general solutions for this equation of motion, where the angular solution is given in terms of the general spheroidal function, and the radial solution is given in terms of the confluent Heun function. Furthermore, they can be chosen according to the boundary conditions to be imposed.

The study of the asymptotic behaviors of the radial solution led to the quasibound state phenomena. In the vicinity of the exterior event horizon, the radial solution diverges by reaching a maximum value, which indicates that the scalar particles cross into the black hole. On the other hand, far from the black hole, at the asymptotic infinity, the radial solution tends to zero, which means that the probability of finding some particles there is null.

We obtained the quasispectrum of resonant frequencies for charged massive scalar particles propagating in the Kerr-Newman black hole spacetime. Furthermore, we analyzed the particular cases of the Kerr and Schwarzschild backgrounds. This become possible by imposing two boundary conditions according to the VBK approach \cite{AnnPhys.373.28,PhysRevD.104.024035} developed to study the quasibound states, which has been successfully used in other scenarios, as for example, Dirac fields in Schwarzschild and Reissner-Nordstr\"{o}m black hole backgrounds \cite{ModPhysLettA.34.1950323}, as well as scalar fields in a rotating linear dilaton black hole \cite{PhysRevD.94.084040}, in a linear dilaton black hole within the Einstein-Maxwell-Dilaton gravity \cite{AIPConfProc.2183.020005}, in the five-dimensional Lovelock black hole spacetime \cite{AnnPhys.433.168583}, and in the Schwarzschild-Anti-de Sitter black hole with an f(R) global monopole \cite{EurPhysJC.81.1143}. We compared our results with the ones known in the literature, and then we conclude that this new method generalizes the numerical approaches used by some authors \cite{PhysRevD.76.084001,ProgTheorPhys.112.983}.

We have found a set of frequencies that are solutions of a characteristic resonant equation, which was derived from the polynomial condition of the confluent Heun functions. The physically acceptable solutions are obtained by analyzing the (asymptotic) behavior of the radial solution related to these resonant frequencies. From this spectrum, we also obtained both the radial and angular wave eigenfunctions.

Finally, we have applied the model to an actual black hole, the supermassive one situated at the center of the M87 galaxy, computing the characteristic time of decay of ultra-light bosonic particles around the black hole, in the relativistic approximation. We verified thus that the magnitude of this time reflects the stability of the bosonic particles around the compact object, which could constitute, for instance, the cold dark matter situated in this region (see Refs.~\cite{ETH3,Lee} and references therein).
%
%
%
%
%
\section*{Data availability}
The data that support the findings of this study are available from the corresponding author upon reasonable request.
%
%
\begin{acknowledgments}
H.S.V. is funded by the Alexander von Humboldt-Stiftung/Foundation (Grant No. 1209836). This study was financed in part by the Coordena\c c\~{a}o de Aperfei\c coamento de Pessoal de N\'{i}vel Superior - Brasil (CAPES) - Finance Code 001. V.B.B. is partially supported by the Conselho Nacional de Desenvolvimento Cient\'{i}fico e Tecnol\'{o}gico (CNPq) through the Research Project No. 307211/2020-7. C.R.M. is partially supported by the CNPq through the Research Project No. 308168/2021-6, and by the Funda\c{c}\~{a}o Cearense de Apoio ao Desenvolvimento Cient\'{\i}fico e Tecnol\'{o}gico (FUNCAP) under the grant PRONEM PNE-0112-00085.01.00/16. The authors would like to thank Dr. Kyriakos Destounis for the very fruitful discussions.
\end{acknowledgments}
%
%

%
%
\end{document}